\documentclass{jfm}
\usepackage{graphicx}
\usepackage{epstopdf, epsfig}

\usepackage[colorinlistoftodos]{todonotes}
\setuptodonotes{inline}
\usepackage{amsmath}

\usepackage{url}

\newcommand{\Rayleigh}{\mathrm{Ra}}
\newcommand{\Rac}{\mathrm{Ra}_{c}}

\newcommand{\Prandtl}{\mathrm{Pr}}
\newcommand{\PrandtlM}{\mathrm{Pm}}
\newcommand{\scrR}{\mathcal{R}}
\newcommand{\scrP}{\mathcal{P}}
\newcommand{\scrS}{\mathcal{S}}
\newcommand{\scrH}{\mathcal{H}}
\newcommand{\scrN}{\mathcal{N}}
\newcommand{\erf}{\mathrm{erf}}

\renewcommand{\vec}{\boldsymbol}

\newcommand{\grad}{\vec{\nabla}}
\newcommand{\lap}{\nabla^2}

\shorttitle{Linear Moist Convection}
\shortauthor{J.\ S.\ Oishi and B.\ P.\ Brown}

\title{On the Linear Stability of Partially and Fully Saturated Atmospheres to Moist Convection}

\author{Jeffrey S. Oishi\aff{1}
  \corresp{\email{jeff.oishi@unh.edu}},
 \and Benjamin P. Brown\aff{2}}

\affiliation{\aff{1}Department of Mechanical Engineering, University of New Hampshire, Durham, NH 03824, USA
\aff{2}Department of Astrophysical and Planetary Sciences, University of
Colorado, Boulder, CO 80309, USA}

\begin{document}

\maketitle

\begin{abstract}
We present a linear analysis of a minimal model of moist convection under a variety of atmospheric conditions. The stationary solutions that we analyze include both fully saturated and partially unsaturated atmospheres in both unconditionally and conditionally unstable cases. We find that all of the solutions we consider are linearly unstable via exchange of stability when sufficiently driven. The critical Rayleigh numbers vary by over an order of magnitude between unconditionally unstable and conditionally unstable atmospheres. The unsaturated atmospheres are notable for the presence of linear gravity wave-like oscillations even in unstable conditions. We study their eigenfunction structure and find that the buoyancy and moisture perturbations are anticorrelated in $z$, such that regions of negative buoyancy have positive moisture content. We suggest that these features in unsaturated atmospheres may explain the phenomenon of gravity wave shedding by moist convective plumes.
\end{abstract}

%\listoftodos
%\todo{Stability figures: Ra vs kx}
%\todo{Do we need to write up an appendix on G-scaling of VPT19}
%\todo{spectra for saturated/unsaturated}
%\todo{NLBVPs for this system are hard}
%\todo{Ben: saturated atmospheres at lower $\Rayleigh$ to match VPT19}
%\todo{Jeff: Get unsaturated matched Chebyshev code working}
%\todo{Run some eigenfunctions as ICs of IVPs, make sure they roughly agree with growth rates (Or even start from noise at low enough $\Rayleigh$ to get agreeement with most unstable mode)}
%\todo{Write up IVP dynamics: moist plumes vs dry rolls and their competition below the wedge}
%\todo{Try EVP of relaxed atmosphere where we see waves in saturated case, since we haven't seen any waves in the unrelaxed background in those cases}

\section{Introduction}
Moist convection is extremely important in the Earth's climate system; it also may affect the dynamics of Jupiter's atmosphere and those of Jovian exoplanets. 
In the Earth's atmosphere, this is caused by the latent heat relased by condensation of water. 
The heat so released can render stable atmospheric motions unstable, leading to extremely complex dynamics.
Moist convection represents a serious challenge to climate modeling and weather forecasting: the phase change occuring on scales of millimeters can affect dynamics on kilometer scales \citep{emanuelAtmosphericConvection1994,stevensATMOSPHERICMOISTCONVECTION2005}.
This process leads to the formation of highly organized structures of clouds, thunderstorms, and dry subsiding motions.
In turn, these structures modify the global scale circulation and play a crucial role in the hydrological cycle: it is thus crucial for accurate prediction that we have a detailed understanding of the self-organization of moist atmospheres.

Even without moisture, convection is remarkably complex. Significant progress in our understanding of dry convection has come from detailed study of simplified models, particularly the celebrated Rayleigh-Benard model. 
Because of its simplicity, Rayleigh-Benard convection can be studied in exquisite theoretical detail via both computational and analytical techniques. These theoretical predictions can be compared to laboratory experiment, providing reproducible solutions that provide considerable insight into the phenomenology of the kinds of highly turbulent convection found in numerous natural and built environments.

However, moisture is in some sense a ``singular perturbation'' to that model; without some explicit treatment of condensation, Rayleigh-Benard solutions provide a very poor approximation to atmospheric dynamics, even on shallow scales where background density variations can be neglected. 
In order to better assess the effects of moisture on convective motions, a number of simplified models have been developed that combine the key advantages of Rayleigh-Benard with an explicit (if parameterized) treatment of moisture.
Some of these models include the Bretherton-Pauluis-Schumacher (BPS) model, detailed in  \cite{brethertonTheoryNonprecipitatingMoist1987}, \cite{pauluisIdealizedMoistRayleighBenard2010}, and \cite{schumacherBuoyancyStatisticsMoist2010}, and the Fast Autoconversion and Rain Evaporation (FARE) model, detailed in \cite{hernandez-duenasMinimalModelsPrecipitating2013}, \cite{dengTropicalCyclogenesisVertical2012}, and \cite{hernandez-duenasStabilityInstabilityCriteria2015}.
All of these models isolate the dynamics arising from the coupling of moisture and convection, dramatically simplifying the complex treatments used in cloud-resolving large eddy simulations \cite[e.g.,][]{guichardShortReviewNumerical2017}.
Because of this, these models allow detailed analysis and physical interpretation while representing minimal models of the key physics.  Simplified models have been used to study cloud aggregation and dependencies on domain aspect ratios \cite[e.g.,][]{pauluisSelfaggregationCloudsConditionally2011}, the impacts of radiative transport \cite[e.g.,][]{pauluisRadiationImpactsConditionally2013}, and how rotation can drive hurricane-like vortices \cite[e.g.,][]{chienHurricaneLikeVorticesConditionally2022}.  It is the simplicity of these models that gives them explanatory power, allowing them to isolate aspects of the full physics present in moist atmospheric convection.

Recently, the Rainy Benard convection (RnBC) model was introduced by \cite[][and hereafter VPT19]{vallisSimpleSystemMoist2019a} as a very simple model of moist convection in the limit that precipitation occurs immediately upon condensation. 
This model offers a highly idealized system that can be studied carefully while retaining the most important element of moisture: a rapid phase transition that releases latent heat.
These three simple models (BPS, FARE, and RnBC) all capture the crucial latent heat release due to condensation and thus allow moisture to drive convective motions.  
They differ in their treatments of precipitation: BPS assumes no precipitation, FARE treats both rain and evaporation, while RnBC assumes all condensed water immediately rains out and leaves the system.
Here, we use the RnBC model to study the linear stability and dynamics of moist convective atmospheres. 
We extend the analysis of the stationary ``drizzle'' solution presented in VPT19 by conducting a comprehensive linear stability analysis; we also use the computation of ``drizzle'' solutions to explore the effects of our numerical approximations in the model itself.

The linear analyses of \cite{brethertonTheoryNonprecipitatingMoist1987}  and \cite{hernandez-duenasStabilityInstabilityCriteria2015} are perhaps closest to our study. However, the former considers the limit in which no precipitation occurs, and the latter assumes inviscid dynamics in the absence of diffusive heat transport.

We explore the Rainy Benard system in some detail. All of our work considers both atmospheres that are fully saturated and atmospheres where the lower portion of the atmosphere is unsaturated. In section~\ref{sec:rainy benard model}, we summarize the essential aspects of the Rainy Benard model and the modeling choices adopted here.  In section~\ref{sec:drizzle} we consider aspects of the static ``drizzle'' solutions to this system, defining three essential regimes of atmospheric parameter space: stable atmospheres, conditionally unstable atmospheres, and unconditionally unstable atmospheres.  In section~\ref{sec:linear stability}, we determine the linear stability of solutions in these different regimes of parameter space using generalized eigenvalue problems, finding that the instability occurs directly via exchange of stability and finding that oscillatory waves exist for some atmospheres even in the presence of instability. Finally, in section~\ref{sec:conclusion}, we draw conclusions and make suggestions for future work. 

\section{The Rainy-Benard model}
\label{sec:rainy benard model}
The RnBC model takes standard Rayleigh-Benard convection for an ideal gas \cite[e.g.][]{spiegelBoussinesqApproximationCompressible1960} and adds an equation describing the mixing ratio of water vapor $q$, which we will refer to throughout as the humidity.
Humidity is dynamically coupled to the buoyancy by a term proportional to the condensation rate, the rate at which liquid water departs the system.
The Rainy-Benard model assumes that this condensation happens faster than any relevant dynamical timescale.
We follow VPT19 in denoting the temperature difference from the mean temperature $\delta T = T - T_m$ as $T$. While doing so is commonplace when working with Rayleigh-Benard, it is much more important to note here as this temperature difference is what goes into the simplified Clausius-Clapyron relationship.

The condensation rate is given by 
\begin{equation}
    \label{eq:condensation}
    C = \frac{(q - q_s) \mathcal{H}(q - q_s)}{\tau},
\end{equation}
where $\tau$ is a model parameter and $\mathcal{H}$ is the Heaviside function.
Under this condensation model, as soon as the humidity reaches its saturation value, any amount of supersaturated water is removed on a timescale $\tau$.
In order that the model be consistent with its assumptions, $\tau$ must be the smallest timescale in the system.

VPT19 gave three different non-dimensionalizations for the system but used the diffusive scaling for calculations. We instead choose the \emph{buoyancy} time $\left(H \theta_0/g \Delta T \right)^{1/2}$, the layer depth $d$, the temperature difference $\Delta T$ across the layer, and the saturation specific humidity at $T = 0$.
For the linear calculations at low $\Rayleigh$ presented here, this is not an important choice.
However, in anticipation of future high-$\Rayleigh$ simulation work, we justify this choice as follows.
Convection features two natural timescales by which a non-dimensionalization can be performed, the diffusion time $\tau_d$ and the buoyancy time $\tau_b$. For high-Rayleigh number RnBC, the ordering of these timescales is $\tau \ll \tau_b \ll \tau_d$. 
Using $\tau_d$ as the non-dimensionalization, one arrives at 
\begin{equation}
    \frac{\tau}{\tau_d} \ll \frac{\tau_b}{\tau_d} \ll 1.
\end{equation}
Referring to the dimensionless $\tau$ as $\tau'$, for the diffusion non-dimensionalization $\tau' = \tau/\tau_d$. The ratio of buoyancy to diffusion time scales is given by $\tau_b/\tau_d = (\Rayleigh \Prandtl)^{-1/2}$, and thus we see that if $\Rayleigh$ increases but $\tau'$ is held fixed, the model assumption will eventually become invalid: $\tau'$ will eventually exceed $\tau_b/\tau_d$.

Instead, if we choose the bouyancy non-dimensionalization with $\tau_b$ as the timescale, we have
\begin{equation}
    \frac{\tau}{\tau_b} \ll 1 \ll \frac{\tau_b}{\tau_d};
\end{equation}
here as long as $\tau' = \tau/\tau_b$ is chosen to be less than one, the model assumptions remain valid as $\Rayleigh$ is increased, so long as the Rayleigh number is large enough that $1 \ll (\Rayleigh \Prandtl)^{1/2}$ to begin with.
Thus, the buoyancy non-dimensionalization provides a key simplification and we adopt it here.

With the buoyancy non-dimensionalization, the Rainy-Benard equations are 
\begin{align}
    \frac{\partial \vec{u} }{\partial t} + \grad p -  b \vec{\hat{z}} - \scrR \lap \vec{u} &= - \vec{u}\cdot\grad \vec{u} \label{eq:NS}\\    
    \frac{\partial b}{\partial t} - \scrP \lap b &= -\vec{u} \cdot \grad b + \gamma C \label{eq:buoyancy}\\  
    \frac{\partial q}{\partial t} - \scrS \lap q &=-\vec{u} \cdot \grad q - C\label{eq:humidity}\\
    \grad \cdot \vec{u} &= 0 \label{eq:continuity}.
\end{align}
The condensation rate $C$ remains as in equation~(\ref{eq:condensation}), though we drop the prime on the dimensionless $\tau$ in what follows.

We replace the Heaviside function in $C$ with a smooth approximation,
\begin{equation}
    \scrH(A) = \frac{1}{2}\left(1 + \erf(k A)\right)
    \label{eq:smooth Heaviside}
\end{equation}
where $k$ controls the slope of the transition (and hence the width of the transition region). We note in passing that this choice is motivated by the fact that $\erf$ provides better convergence properties than $\tanh$ for a given slope parameter $k$.

As in VPT19, the simplified Clausius–Clapeyron relation is
\begin{equation}
  \label{eq:Clausius-Clapeyron}
    q_s = \exp{(\alpha T)}
\end{equation}
and the temperature is related to the buoyancy via
\begin{equation}
T = b - \beta z,
\end{equation}
where $\beta = d g/\Delta T c_p$ is the ratio of the adiabatic gradient to the overall temperature gradient across the layer.
This parameter formally exists in Rayleigh-Benard convection when it is derived from an ideal gas in a shallow layer \citep{spiegelBoussinesqApproximationCompressible1960}, however it has no dynamical effect, as the background buoyancy gradient is absorbed into the pressure gradient term.
However, in this system, the temperature differential $T$ appears in equation~
(\ref{eq:Clausius-Clapeyron}).
We will discuss the implications of this in section~\ref{sec:drizzle}.

Finally, the dimensionless parameters are
\begin{equation}
    \scrR = \left(\frac{\Prandtl}{\Rayleigh}\right)^{1/2}, \quad
    \scrP = \frac{1}{(\Rayleigh \Prandtl)^{1/2}}, \quad
    \scrS = \frac{1}{(\Rayleigh \PrandtlM)^{1/2}},
\end{equation}
where $\Prandtl = \nu/\kappa$ and $\Rayleigh = g \Delta T d^3/T_m \nu \kappa$ are the Prandtl and Rayleigh numbers, respectively.

For all eigenvalue and non-linear boundary value problems presented here, we use the Dedalus framework \citep{burnsDedalusFlexibleFramework2020}. 
We use Legendre polynomials for discretization in $z$, typically using $n_z = 32$--$128$ spectral modes, and a generalized tau formulation for the boundary conditions \citep{burnsCornerCasesTau2024}.
We use the eigentools package \citep{oishiEigentoolsPythonPackage2021a} to ensure that our eigenvalue solutions are well resolved using mode rejection performed using two different basis functions at the same resolution, effecting a speedup by a factor of approximately $2.2$ over the traditional comparison of a solution with $3N/2$ modes.

\section{Drizzle solutions}
\label{sec:drizzle}
VPT19 refer to the equilibrium state for this system as the ``drizzle'' solution. For convenience we briefly restate some results from that work to provide context for the novel results in subsequent sections; we also provide a library of reference states that we will refer to throughout our exploration of the linear dynamics. The drizzle state is a hydrostatic solution with constant precipitation and moisture replaced via diffusion across the lower boundary. Exact solutions can be obtained in terms of the principal branch of the Lambert-W function.
In the case with a saturated lower boundary, this is a straightforward procedure; for cases with an unsaturated lower boundary the process is slightly more involved.
We expand on the details in the latter case from those provided in \cite{vallisSimpleSystemMoist2019a} in the appendix. We note that all the states we consider have saturated conditions in the upper part of the domain, which is uncommon in Earth's atmosphere. However, the stability and linear dynamics of these states provide a simple and interpretable framework for understanding the differences between moist convection and its dry counterparts. The RnBC model itself is capable of more realistic fully unsaturated initial conditions, though we defer detailed investiagions of such states to future work. 

We will also use the analytic drizzle solutions to benchmark numerical approximations. This is particularly important when interpreting results from non-linear simulations in which numerical approximations to the Heaviside function must be made. Furthermore, the requirement that $\tau$ must be the smallest timescale leads to stringent constraints on the timestep, which is usually set by CFL conditions applied to the smallest scales of the flow. By studying the convergence of non-linear boundary value problems (NLBVP) to the analytic solution while varying the numerical model parameters $k$ and $\tau$, we provide guidelines for the computational resources required for accurate simulations.
%in the saturated portion of the atmosphere, matched onto linear profiles in any unsaturated portions of the atmosphere (typically at the bottom of the atmosphere, given the scaling of saturation with temperature and falling temperature profiles with height).

The drizzle solution can be expressed rather simply by making use of the moist static energy $m = b + \gamma q$. In the static limit ($\mathbf{u} = 0$), the moist static energy equation can be derived from adding equation~(\ref{eq:buoyancy}) to $\gamma$ times equation~(\ref{eq:humidity}) to arrive at
\begin{equation}
\nabla^2 m = \nabla^2(b + \gamma q) = 0.
\label{eq:steady_m}
\end{equation}
This has a simple solution in $z$,
\begin{equation}
  \label{eq:m_drizzle}
  m(z) = P + Q z
\end{equation}
with
\begin{align}
P &= b_1 + \gamma q_1, \\
Q & = (b_2 - b_1) + \gamma (q_2-q_1)
\end{align}
where $b_{1,2}$ and $q_{1,2}$ are the values at the boundaries.  

\subsection{Saturated atmospheres}
\label{sec:saturated_atmospheres}
The profile of $T$ in the saturated portion of the atmosphere is given by
\begin{equation}
  \label{eq:T_drizzle}
T = C(z) - \frac{W\Big(\alpha \gamma \exp{(\alpha C(z))}\Big)}{\alpha},
\end{equation}
with
\begin{equation}
  \label{eq:bigC}
  C(z) = P + (Q-\beta) z;
\end{equation}
$W$ is the Lambert-W function.

A saturated atmosphere has $q=q_s$ at the bottom and hence $r_h = 1$ everywhere. We take boundary values $b_1 = 0$, $b_2 = \beta + \Delta T$, $q_1 = 1$, $q_2 = \exp{(\alpha \Delta T)}$ and $\Delta T = -1$.  These lead to the simplifications:
\begin{align}
\label{eq:P_saturated} P &= \gamma\\
Q & = \beta - 1 + \gamma \Big(\exp{(-\alpha)}-1\Big), \\
\label{eq:C_saturated} C(z) & = \gamma \left( 1 + \Big(\big(\exp{(-\alpha)}-1\big) -1\Big) z\right)
\end{align}
To obtain a static solution, one solves for $m(z)$ and $T(z)$, sets $q(z)=q_s=\exp(\alpha T(z))$ (e.g., relative humidity $r_h=1$) and then obtains $b(z) = m(z) - \gamma q(z)$.  From equations~(\ref{eq:P_saturated}--\ref{eq:C_saturated}), it is clear that $m$ and $b$ will depend on $\beta$, while $T$ (and hence $q$) will not, while all will vary with $\gamma$.  Indeed, this is what we find.

\begin{figure}
    \centering
    \includegraphics[width=\textwidth]{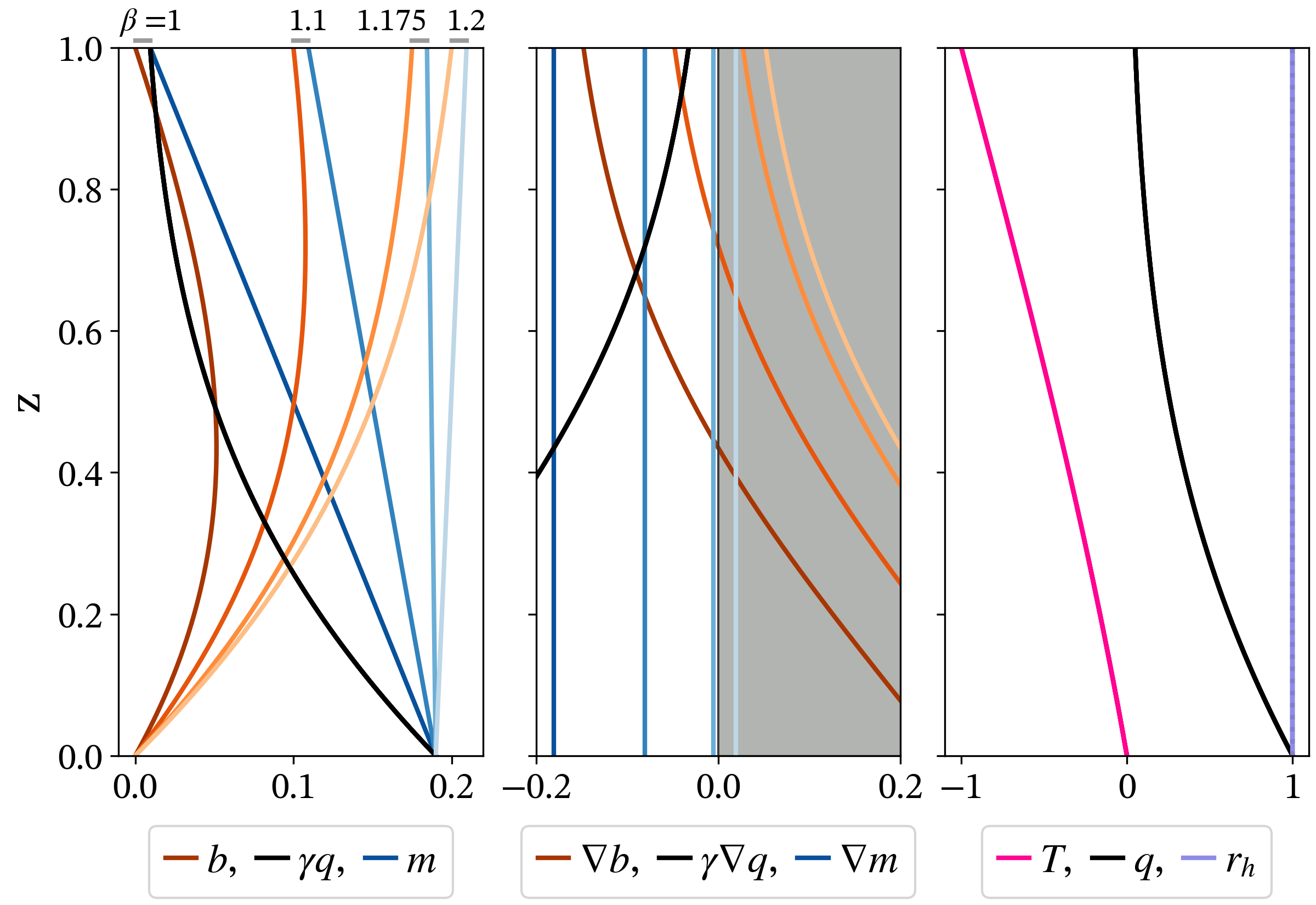}
    \caption{Saturated atmospheres with $\alpha=3$ and $\gamma=0.19$, with varying values of $\beta$ (labelled above corresponding $b$, $m$ in left figure).  Shown at left are profiles of buoyancy $b$, scaled moisture $\gamma q$ and moist static energy $m$.  The profiles of $b$ and $m$ change with $\beta$, gradually increasing the value of the gradient from negative (convectively unstable) to positive (convectively stable, grey region of panel).  These gradients are shown in the middle panel.  The profile of $q$ is independent of $\beta$, as it follows $q_s$ which depends on $T$, which is itself independent of $\beta$ (right panel).  The temperature profile is given by a Lambert-W function and is not linear.  The relative humidity is $r_h=1$ everywhere in the saturated atmosphere.
    \label{fig:saturated_atmospheres}}
\end{figure}

A selection of different saturated atmospheres, all at $\alpha=3$ and $\gamma=0.19$ (Earth-appropriate values) are shown in Figure~\ref{fig:saturated_atmospheres}.  The profiles of $b$ and $m$ change as $\beta$ varies, but $T$ is independent of $\beta$ and consequently so are both the saturation humidity $q_s$ and humidity $q$.  The profile of $T$ is not linear with height, in contrast to regular Rayleigh-Benard equilbria, though the profile of $m$ is.
The importance of $\beta$ is related directly to that fact. When Rayleigh-Benard convection occurs in thin layers of compressible ideal gas, the only function $\beta$ has is to delinate the ideal stability (that is, in the absence of diffusive effects): any $\beta \ge 1$ is stable. However, this is because the background gradient $\partial_z T_0 = -1$ due to the fact that the steady state solution is one in which $\nabla^2 T = 0$, and buoyancy and temperature are equivalent in Rayleigh-Benard. 
The case is different for RnBC, where buoyancy is coupled to both temperature and moisture.  With an additional physical quantity comes an additional dimensionless parameter: $\beta$.
As a direct result, RnBC drizzle solutions have ideal stability limits somewhat different from $\beta = 1$; this will be made clear in the following discussion.

The ideal stability of the atmosphere can be deduced from the gradients $\nabla m$, $\nabla b$, and $\nabla q$.  The gradient $\nabla m$ is single-valued and when negative the atmosphere is unstable to moist convection (``moist unstable''). The buoyancy gradient $\nabla b$ changes with height, typically being smallest (or most negative) near the top of the atmosphere.  In all atmospheres shown in figure~\ref{fig:saturated_atmospheres}, $\nabla b$ is positive in the lower portion of the atmosphere, and in some it is negative in the upper portion;  atmospheres with $\nabla b< 0$ somewhere are \emph{unconditionally unstable}.  For some values of $\beta$ (e.g., $\beta = 1.175$), $\nabla b$ is positive everywhere, these are \emph{conditionally unstable} atmospheres.

\begin{figure}
    \centering
    \includegraphics[width=\textwidth]{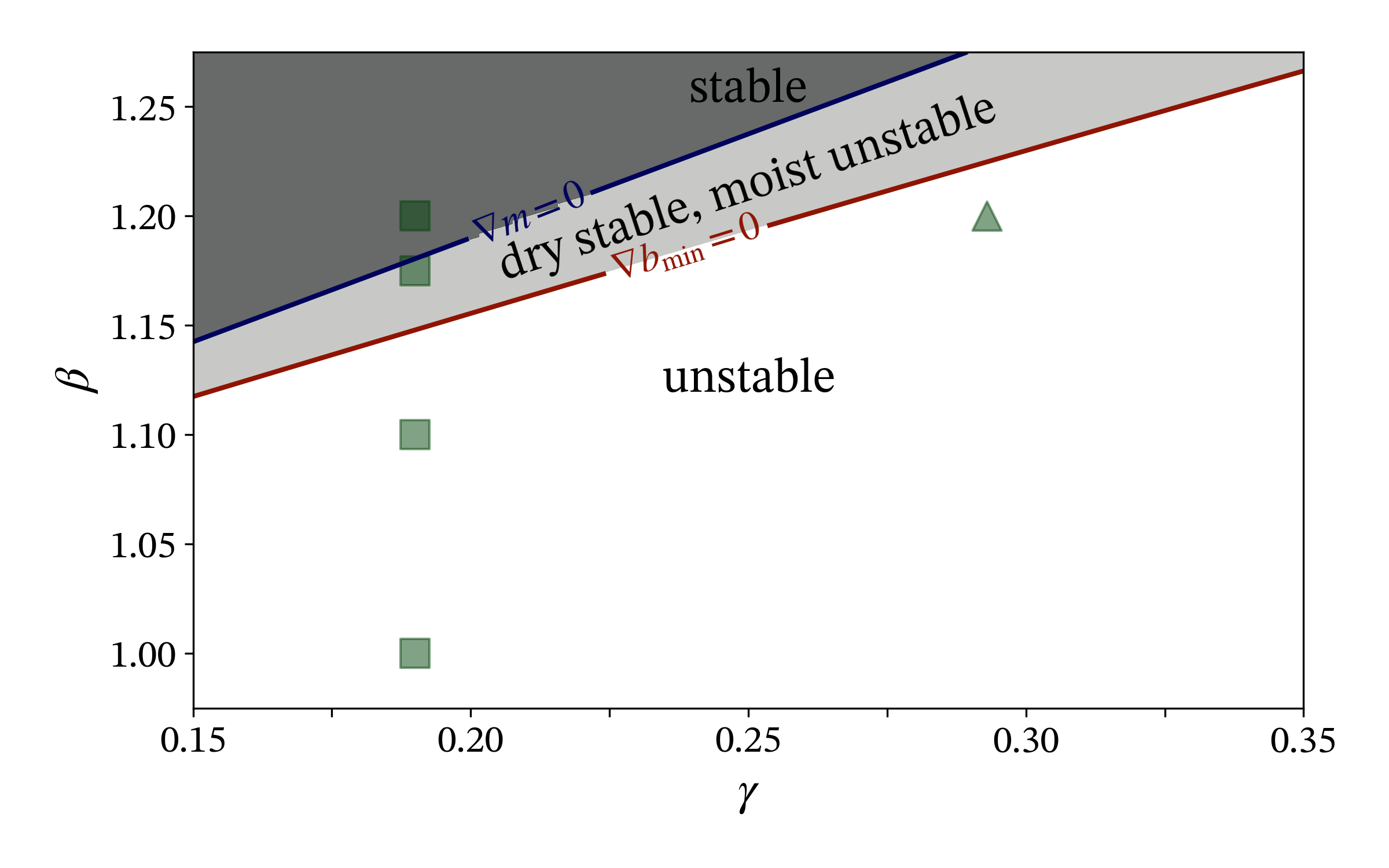}
    \caption{Ideal stability of saturated Rainy-Benard atmospheres.  Shown are the boundaries to moist instability ($\nabla m = 0$) and to dry instability ($\nabla b_{min} =0$, with this the smallest or most negative value of $\nabla b$).  At fixed $\gamma$, as $\beta$ increases, the atmosphere goes from fully unstable to dry stable but moist unstable (light grey wedge) before becoming fully stable (dark grey region).  Squares show points from Figure~\ref{fig:saturated_atmospheres}, and the triangle shows the scaled atmosphere from section 6 of \cite[][VPT19, and see appendix~\ref{sec:VPT19 correction}]{vallisSimpleSystemMoist2019a}.
    \label{fig:saturated_ideal_stability}}
\end{figure}

The results shown in figure~\ref{fig:saturated_atmospheres} for $\gamma=0.19$ hold more broadly as $\gamma$ varies, though the details for which $\beta$ are stable, conditionally unstable, or unconditionally unstable depend on $\gamma$.
The ideal stability boundaries for saturated atmospheres at many $\gamma$ are shown in figure~\ref{fig:saturated_ideal_stability}.
The region of conditional instability, where the atmosphere is unstable to moist instability ($\nabla m < 0$), but stable to dry stability ($\nabla b > 0$ at all $z$), defines a limited wedge in parameter space.
The range of $\beta$ where this occurs is $\gamma$-dependent, as shown in figure~\ref{fig:saturated_ideal_stability} with a light grey wedge.  
This wedge converges at $\gamma=0$ to $\beta=1$ (as expected for Rayleigh-Benard convection) and widens as $\gamma$ increases.
This wedge of conditional instability is an interesting region where dynamics are likely to be distinctly different from classic thermal Rayleigh-Benard convection.
Here, for fully-saturated atmospheres, we choose $\gamma=0.19$, $\beta=1.175$ as our representative conditionally unstable atmosphere, and $\gamma=0.19$, $\beta=1.1$ as our representative unconditionally unstable atmosphere.

\subsection{Unsaturated atmospheres}
\label{sec:unsaturated_atmospheres}
In unsaturated atmospheres, the lower boundary is at a moisture value below the saturation point of the atmosphere at that height.  The humidity, buoyancy and moist static energy $m$ profiles are all linear in the unsaturated portion of the atmosphere \citep{vallisSimpleSystemMoist2019a}.  While the humidity $q$ generally declines (linearly) with height, the saturation humidity $q_s$ declines exponentially with decreasing temperature.  At a critical height $z=z_c$ and a critical temperature $T(z_c) = T_c$ the atmosphere becomes saturated.  From this height and above, the solution follows the saturated solutions from Section~\ref{sec:saturated_atmospheres}, but using the values at $z=z_c$ as the basal values for the solution.

\begin{figure}
    \centering
    \includegraphics[width=0.49\textwidth]{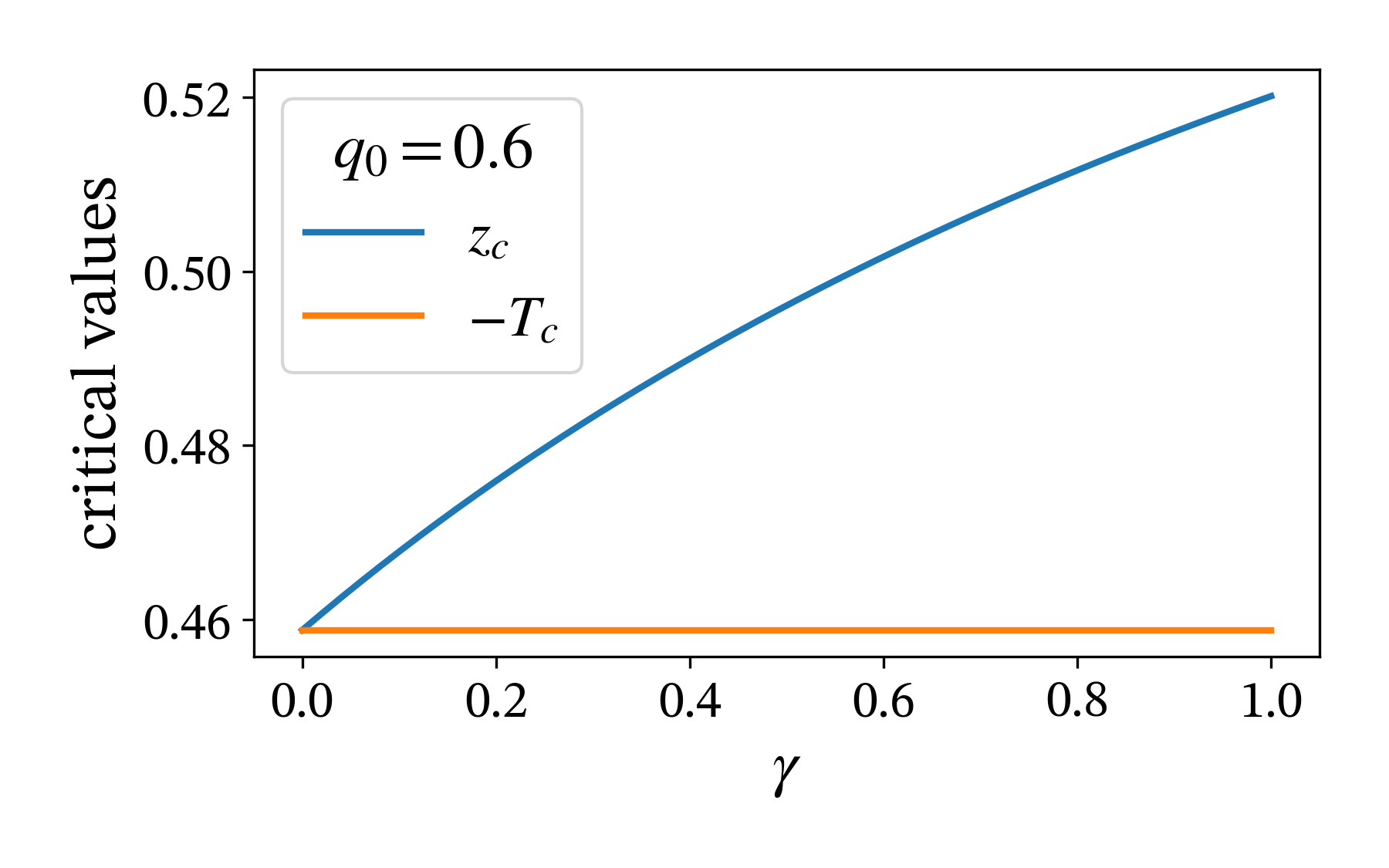}
    \includegraphics[width=0.49\textwidth]{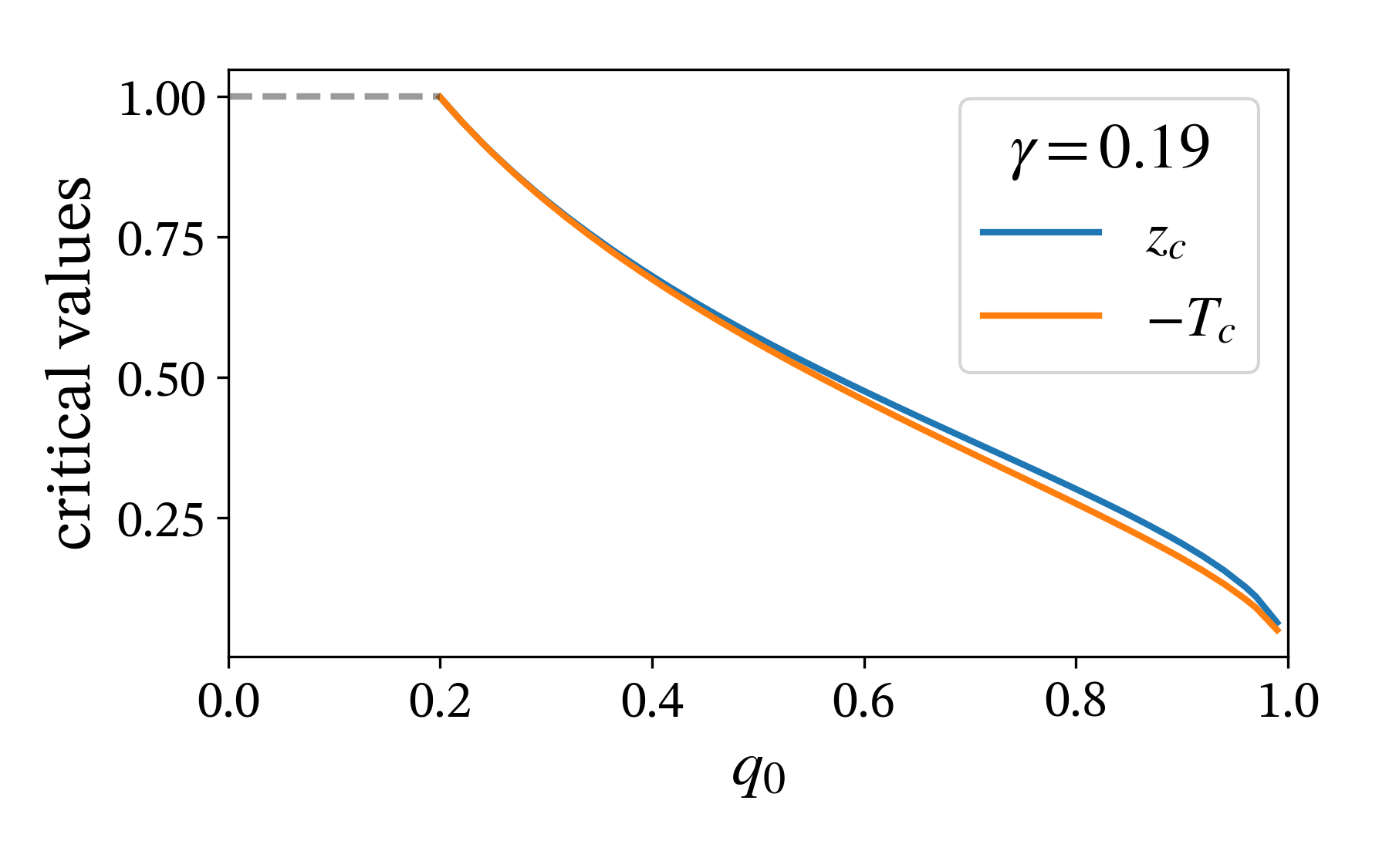}
    \caption{(left) Critical height $z_c$ at which an atmosphere with an unsaturated lower boundary becomes saturated and the temperature $T_c$ at that height, both as functions of $\gamma$. These solutions are at $q_0 = 0.6$.  Note $z_c$ varies with $\gamma$ while $T_c$ is independent of $\gamma$. (right) Critical height $z_c$  and temperature $T_c$ at which an atmosphere with an unsaturated lower boundary becomes saturated as a function of relative humidity at the lower boundary $q_0 = q(z=0)$.  Here $\gamma=0.19$, and both $z_c$ and $T_c$ depend on $q_0$.  We find no dependence of $z_c$ or $T_c$ on $\beta$.}
    \label{fig:zc_tc_vs_gamma_and_q0}
\end{figure}

The dependence of $z_c$ and $T_c$ on atmospheric parameters for $\alpha=3$ is shown in Figure~\ref{fig:zc_tc_vs_gamma_and_q0}.  We find that $z_c$ depends on $\gamma$, while $T_c$ is independent of $\gamma$.  Both $z_c$ and $T_c$ depend on $q_0 = q(z=0)$, the relative humidity at the lower boundary, and these behave sensibly in the limit of a saturated atmosphere ($q_0 = 1$).  The variations in $\gamma$, though significant, are much smaller than the variations in $q_0$; when $\gamma$ increases, so does $z_c$, while $T_c$ remains unchanged.  Neither $z_c$ nor $T_c$ shows any dependence on $\beta$ in the ranges we have tested $\beta=[1,1.2]$ (not shown).  Below a critical $q_0 \approx 0.2$ there are no solutions, as $z_c > 1$ violates our assumptions; this critical $q_0$ appears to be independent of $\gamma$.  

\begin{figure}
    \centering
    \includegraphics[width=\textwidth]{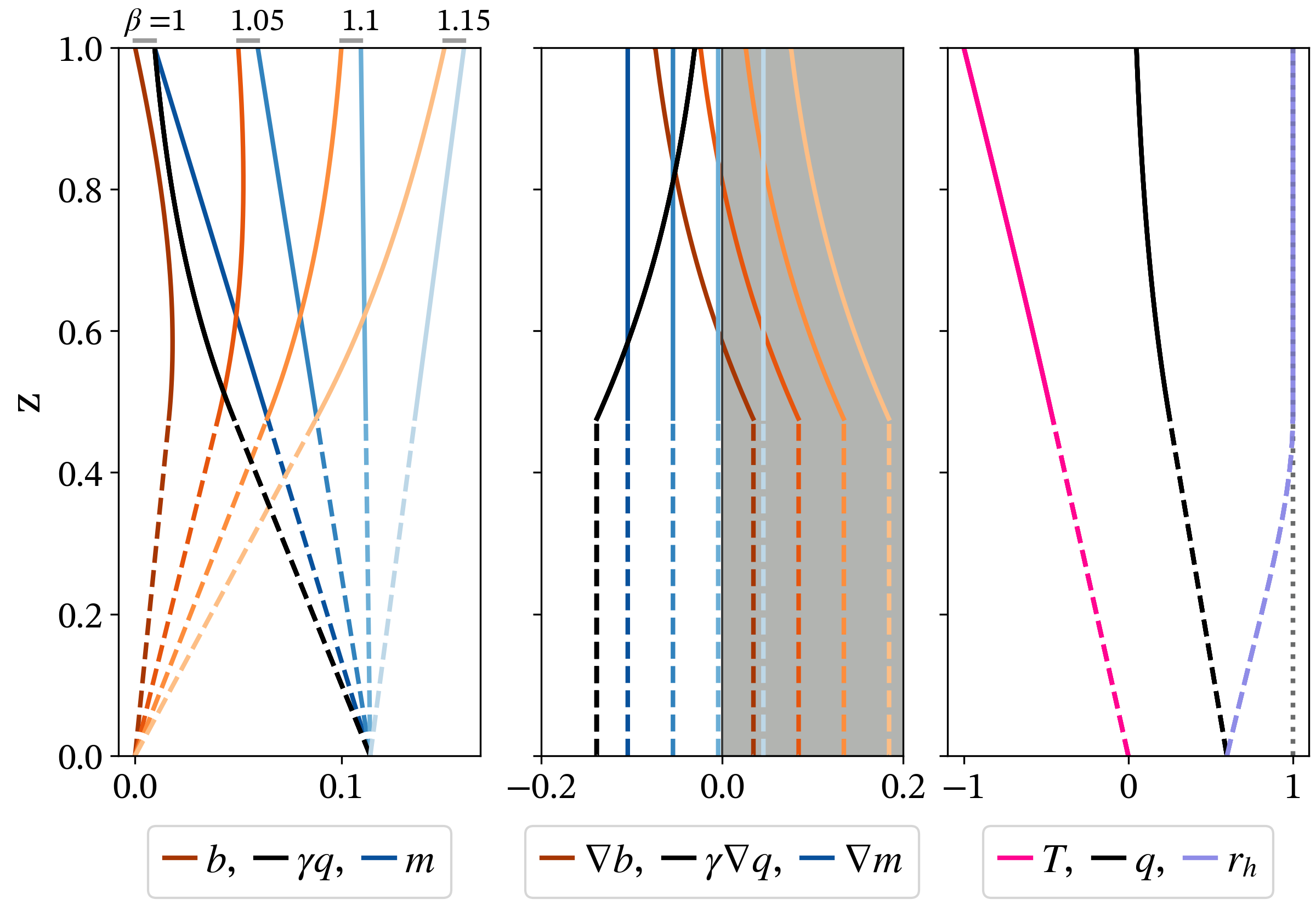}
    \caption{Unsaturated atmospheres with $\alpha=3$ and $\gamma=0.19$, and with $q(z=0)=0.6$, at varying values of $\beta$ (labelled above corresponding $b$, $m$ in left figure).  In this atmosphere, $z_c \approx 0.475$ and $T_c \approx -0.459$, independent of $\beta$. Shown at left are profiles of buoyancy $b$, scaled humidity $\gamma q$ and moist static energy $m$.  Below $z_c$ (dashed lines) the profiles are linear, while above $z_c$ (solid lines) $q(z)$ and $b(z)$ have nonlinear structure.  The profiles of $b$ and $m$ change with $\beta$, gradually increasing the value of the gradient from negative (convectively unstable) to positive (convectively stable, grey region of panel).  These gradients are shown in the middle panel.  The profile of $q$ is independent of $\beta$; this is true both below $z_c$ and above, where it follows $q_s$ which depends on $T$, which is itself independent of $\beta$ (right panel).  The temperature profile is given by a Lambert-W function and is not linear above $z_c$.  The relative humidity is $r_h<1$ below $z_c$ and $r_h = 1$ for $z\geq z_c$.
    \label{fig:unsaturated_atmospheres}}
\end{figure}

A selection of unsaturated atmospheres, all at $\alpha=3$ and $\gamma=0.19$ with $q(z=0)=q_0=0.6$, are shown in Figure~\ref{fig:unsaturated_atmospheres}.  The full details on constructing these piecewise solutions to unsaturated atmospheres are in Appendix~\ref{sec:piecewise}. 
In this atmosphere, $z_c \approx 0.475$ and $T_c \approx -0.459$ for all values of $\beta$.
The saturated portion of the atmosphere above $z=z_c$ shows similar nonlinearity in $q(z)$ and $b(z)$ as in Figure~\ref{fig:saturated_atmospheres}, while the unsaturated portions below $z=z_c$ are linear and the profiles of $m(z)$ are linear throughout.  The profiles of $T(z)$ are nonlinear above $z=z_c$, though this is difficult to discern in the plot.  As with the saturated atmospheres, $b(z)$ and $m(z)$ depend on $\beta$, while $q(z)$ and $T(z)$ do not.  The character of the gradients is similar to what was found for saturated atmospheres, and as before, atmospheres can be either fully unstable (e.g., $\beta=1$), moist unstable but dry stable (e.g., $\beta = 1.1$), or fully stable (e.g., $\beta=1.15$).  In all cases, the relative humidity $r_h$ is less than one below $z=z_c$ and $r_h=1$ above that critical point.

\begin{figure}
    \centering
    \includegraphics[width=\textwidth]{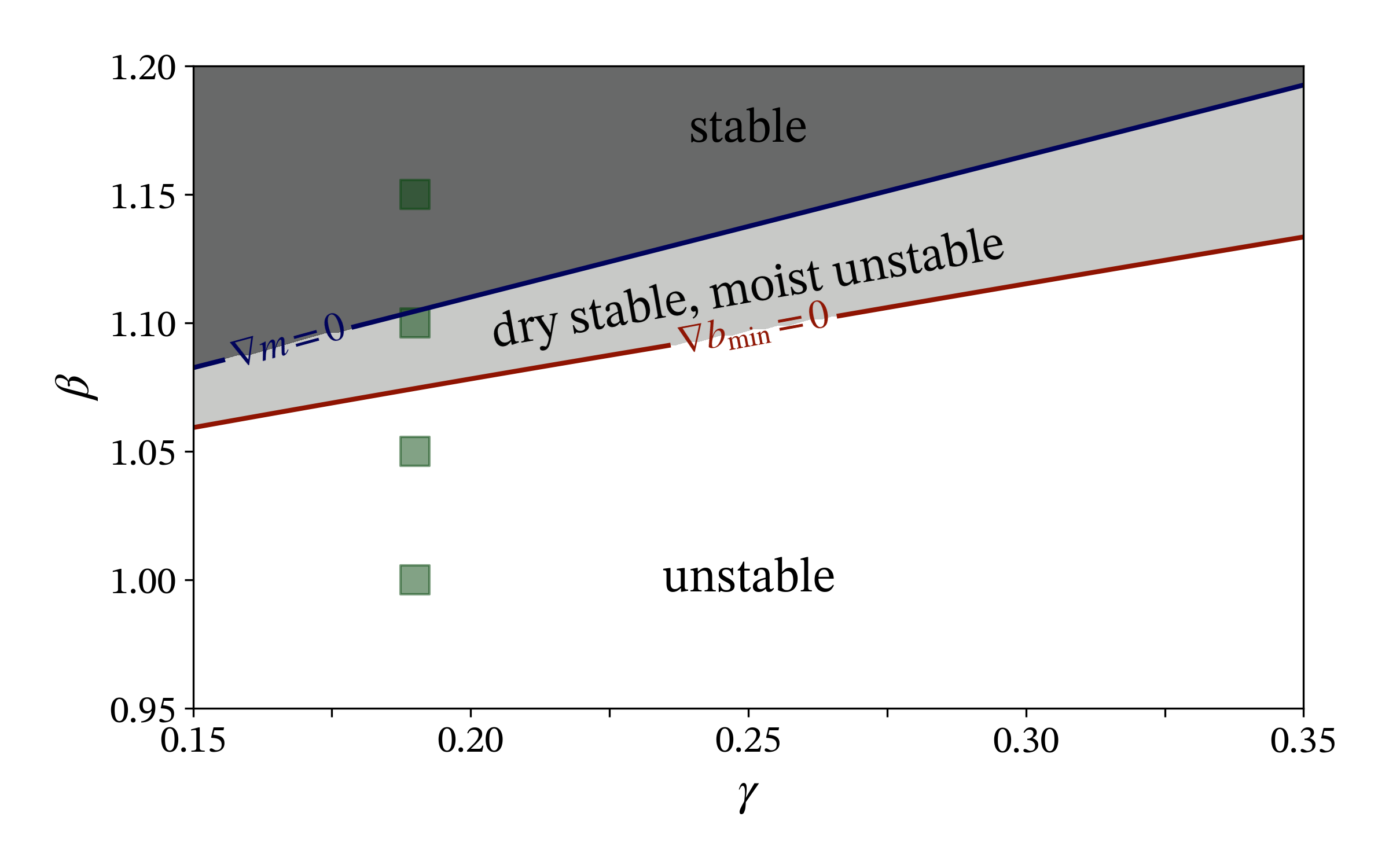}
    \caption{Ideal stability of unsaturated ($q(z=0)=0.6$) Rainy-Benard atmospheres.  Shown are the boundaries to moist instability ($\nabla m = 0$) and to dry instability ($\partial_z b_{min} =0$, with this the smallest or most negative value of $\partial_z b$).  At fixed $\gamma$, as $\beta$ increases, the atmosphere goes from fully unstable to dry stable but moist unstable (light grey wedge) before becoming fully stable (dark grey region).  Squares show points from Figure~\ref{fig:unsaturated_atmospheres}.
    \label{fig:unsaturated_ideal_stability}}
\end{figure}

As for the saturated atmosphere, the ideal stability of the unsaturated atmospheres based on the gradients $\nabla m$ and $\nabla b$ is shown in Figure~\ref{fig:unsaturated_ideal_stability}.  Atmospheres can again be either stable, conditionally unstable, or unconditionally unstable.  Generally, at fixed $\gamma$, the unsaturated atmospheres become stable for smaller values of $\beta$ than their saturated counterparts (compare to Figure~\ref{fig:saturated_ideal_stability}).  The interesting wedge of dry stability and moist instability remains present and, as before, this wedge converges at $\gamma=0$, and widens as $\gamma$ increases.  
For unsaturated atmospheres with $q(z=0)=0.6$, we choose $\gamma=0.19$, $\beta =1.1$ as our representative conditionally unstable atmosphere, and $\beta=1.05$ as our representative unconditionally unstable atmosphere.

\subsection{Drizzle solutions via non-linear boundary value problems}

When approaching a system like Rainy-Benard, one might naturally attempt computing equilibrium solutions using numerical tools for the full coupled nonlinear system in equilibrium:
\begin{align}
\scrP \nabla^2 b &= \frac{\gamma}{\tau}(q-q_s)\mathcal{H}(q-q_s), \\
\scrS \nabla^2 q &= -\frac{1}{\tau}(q-q_s)\mathcal{H}(q-q_s), \\
q_s &= \exp{(\alpha (b - \beta z))}
\end{align}
seeking solutions to $b$ and $q$ subject to boundary conditions.
In the remainder of this section, we set $\scrP = \scrS$.
Using the Dedalus non-linear boundary value problem (NLBVP) solver to compute drizzle solutions with saturated and unsaturated lower boundaries, we found them to be difficult to converge.  

While the rest of this manuscript utilizes the analytic atmosphere solutions discussed in Sections~\ref{sec:saturated_atmospheres} and \ref{sec:unsaturated_atmospheres}, here we present some details of NLBVP solutions.  This section will act both as a word of caution and to help isolate the effect of numerical approximations to the Heaviside function on resulting equilibria.

%We have computed drizzle solutions with saturated and unsaturated lower boundaries using the Dedalus non-linear boundary value problem (NLBVP) solver. 
%While both fully saturated atmospheres and those which reach saturation somewhere within the domain have analytical and asymptotic solutions, respectively, by using the NLBVP method we can isolate the effect of numerical approximations to the Heaviside function.

In this section, we assess convergence via relative error in the humidity variable, comparing the NLBVP computed $q(z)$ to the analytic solution $q_A(z)$:
\begin{equation}
    E_q = \frac{\int|q(z) - q_A(z)| dz}{\int|q_A(z)| dz},
\end{equation}
We find that convergence in $E_q$ is similar to convergence in other variables in the system (e.g., buoyancy or relative humidity). 

For saturated atmospheres, where $q(z=0)=1$, $\gamma=0.19$, $\beta=1.1$, the NLBVPs quickly converge to an accurate solution.  
These atmospheres can be solved using two different approaches to the condensation rate.  In the simplest approach, we explicitly set $\scrH(A) = 1$, as the atmosphere is everywhere saturated.  This removes the numerical effects from a finite-width Heaviside function.  The system remains nonlinear via $q_s(z)$. 
Under this approach, sampling in $3 \times 10^{-6} \leq \tau \scrP \leq 10^{-3}$, we find that convergence is independent of resolution above a very low threshold (e.g., we found good solutions from $n_z=8$ to $n_z=1024$), and convergence depends only on $\tau$, with $E_q \propto \tau$.  Using our techniques, we were unable to converge solutions at lower $\tau$ at any resolution (e.g., $\tau \scrP \lesssim 10^{-6}$).

Alternatively, we can take a smooth Heaviside for $\scrH(A)$ as in equation~\ref{eq:smooth Heaviside}.  Now we are able to explicitly test the numerical effects of the approximate condensation rate used in the rest of our work.  Here we sample in both $3 \times 10^{-6} \leq \tau \scrP \leq 10^{-3}$ and in $10^3 \leq k \leq 10^5$.  For these saturated atmospheres, we find essentially no dependence on $k$, and a similar $E_q \propto \tau$ dependence.

The story is different for unsaturated atmospheres, where $q(z=0)<1$.  Here we compute NLBVP solutions with $q(z=0)=0.6$, $\gamma=0.19$, $\beta=1.1$ using $\scrH(A)$ as in equation~\ref{eq:smooth Heaviside} and sample in both $3 \times 10^{-6} \leq \tau \scrP \leq 10^{-3}$ and in $10^3 \leq k \leq 10^5$.  
As $\tau$ decreases, the NLBVP system becomes increasingly difficult to converge, even with increased resolution and irrespective of initial guesses for the solution.
The underlying problem appears to be related to linearization of the NLBVP system during iterative convergence: as discussed in section~\ref{sec:linear stability} (see equations~\ref{eq:AHA} and \ref{eq:AHA O(A1)}), expansion of the smooth Heaviside $\scrH(A)$ includes a term that is very small in the vicinity of the transition.  To solve this problem in our NLBVPs, we restricted our solver from expanding and linearizing $\scrH(A)$ during the interative convergence.  This dramatically improved the ability of the NLBVP solver to converge.  In general, we found fastest convergence by starting at large $k$ with the analytic solution as our guess, and then using continuation techniques to continue in the direction of decreasing $k$ at fixed $\tau$. 

Starting from the analytic solution highlights two important issues. First, NLBVP approaches to finding drizzle solutions remains extremely challenging and analytic techniques remain essential to make progress--even knowing the answer requires care in handling $k$ and $\tau$. Second, these tests are reasonable proxies for nonlinear simulations, as those will typically be updating data from prior time steps and thus will begin close to the analytic solution and then drift slowly in a given timestep.

\begin{figure}
    \centering
    \includegraphics[width=0.45\textwidth]{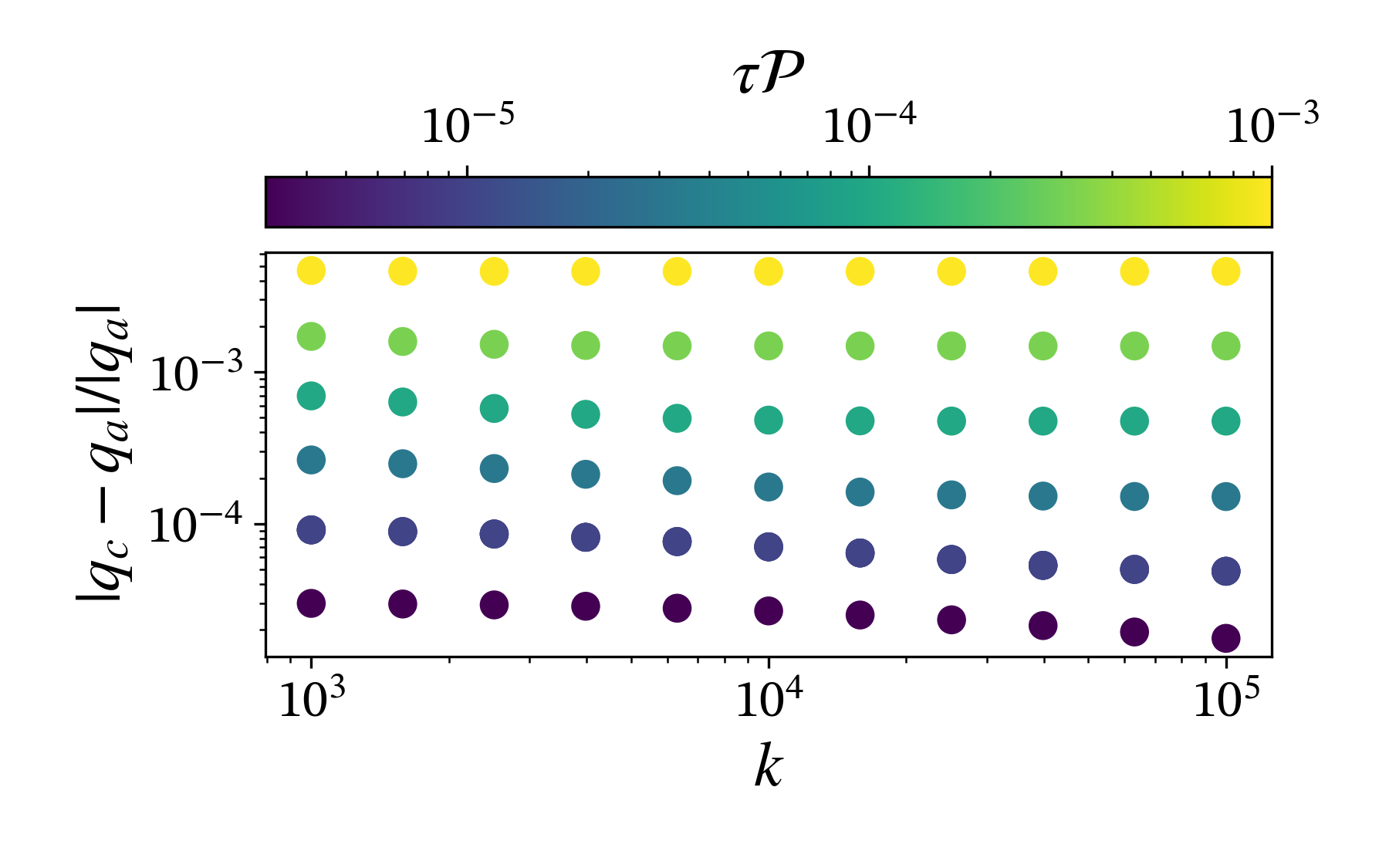}
    \includegraphics[width=0.45\textwidth]{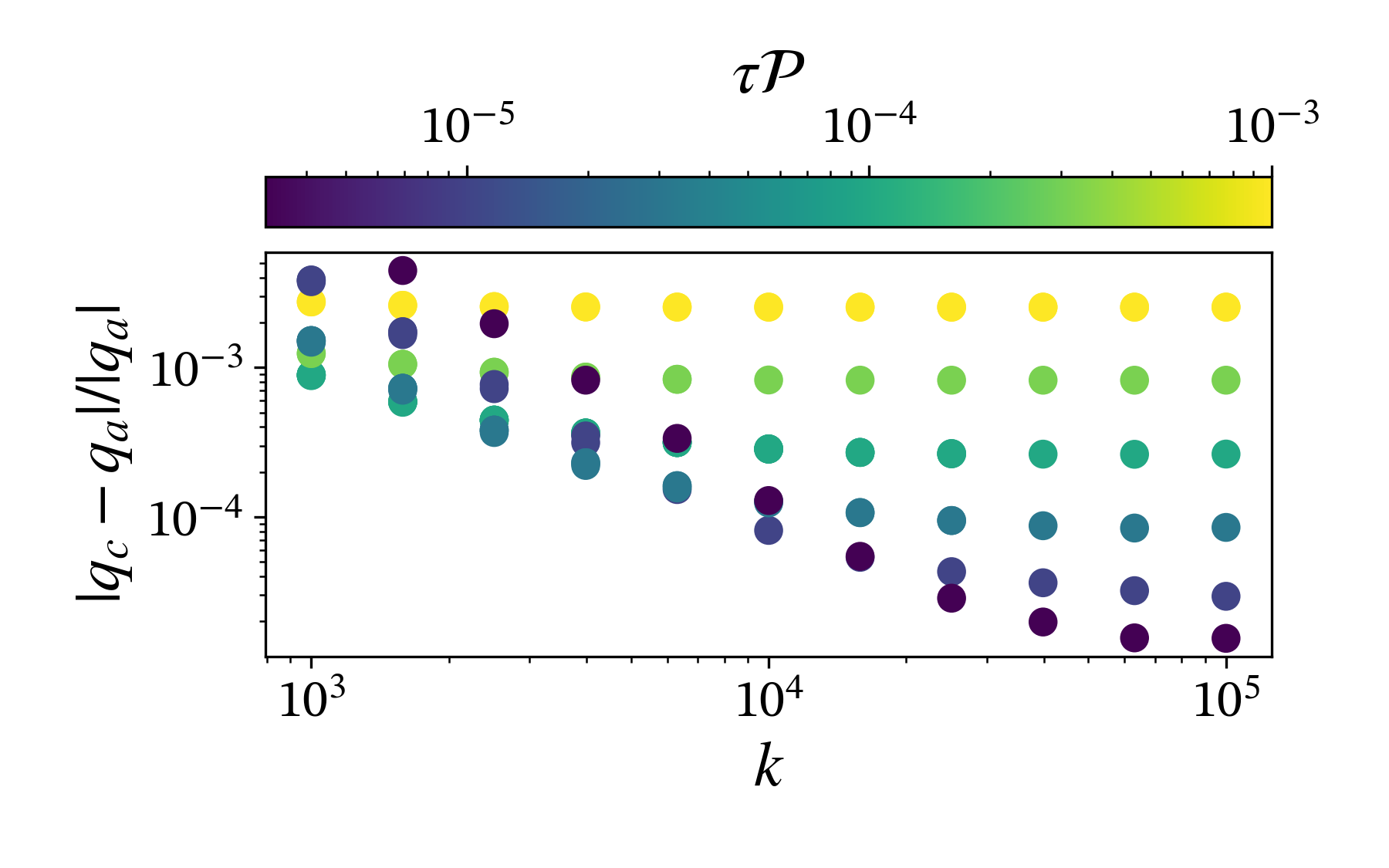}
    \caption{Convergence of NLBVP solutions with $\tau$ and $k$ for saturated (left) and unsaturated (right) atmospheres with $\beta=1.1$.  Here we assess $E_q$ for the computed humidity variable $q_c$, compared to the same from the analytic solution $q_a$.  Saturated atmospheres are straightforward to converge, and the degree of convergence depends nearly only on $\tau$.  For unsaturated atmospheres, a dependence on both $\tau$ and $k$ is visible.
    \label{fig:NLBVP_convergence}}
\end{figure}

The resulting solutions in $\tau$ and $k$ are shown in figure~\ref{fig:NLBVP_convergence} (left).
At large $\tau\scrP =10^{-3}$, the solutions do not depend on $k$.  As $\tau$ decreases, solutions initially depend strongly on $k$.  At low $k$, accuracy decreases as $\tau$ increases for the saturated case (left panel), as expected; however in the unsaturate case (right panel) the accuracy at low $k$ decreases as $\tau$ decreases, somewhat counter-intuitively.  In all systems at high $k$, the solutions approach the analytic solution. 
At fixed $\tau$ and large $k$ we generally find that the accuracy plateaus at some fixed $E_q$.  The plateau values in the unsaturated atmospheres are similar to those found in the fully saturated atmospheres.

Given that a nonlinear simulation will also involve iterated evaluations of these nonlinearities, we expect that these results will set a floor on the overall accuracy, particularly with regard to $\tau$.
A properly self-consistent simulation must choose $\Delta t \ll \tau$, placing strict limits on timestep size. 
There is tension between the weak convergence of NLBVPs with $\tau$ observed here and the desire for high accuracy in nonlinear timestepping simulations.
Meanwhile, nonlinear simulations in unsaturated atmospheres desiring high accuracy will also need to select sufficiently large $k$, but the influence of $k$ is governed by $\tau$. 
We will consider these interrelated issues further in subsequent work.

\section{Linear Stability}
\label{sec:linear stability}
Having established the model equations and the character of their static solutions, we now turn to the main result of this work.
We seek the linear stability of the ``drizzle'' solutions with both saturated and unsaturated lower boundaries, for both conditionally and unconditionally unstable atmospheres. 
To do so, we linearize equations \ref{eq:NS}-\ref{eq:continuity} about four representative drizzle solutions and formulate them as generalized eigenvalue problems.

% Figure~\ref{fig:zc_tc_vs_gamma_and_q0} shows the dependence of $z_c$ and $T_c$ on $\gamma$.
% Importantly, $z_c$ is insensitive to changes in $\beta$, assuming as we do that the boundary conditions on temperature are fixed at $T(z=0) = -1, T(z=1) = 0$.
% This means that once $\gamma$ is fixed by the condensate, $z_c$ and $T_c$ are set entirely by the lower boundary condition on moisture, $q_0 \equiv q(z=0)$. 
% Here, we consider the case where $q_0 = 0.6$.

Linearzation of the RnBC equations is significantly complicated by the Heaviside function in equation~\ref{eq:condensation}. 
In the saturated atmospheres, at leading order, the Heaviside function is $\scrH(z) = 1$ everywhere, and in these atmospheres the effect of $\scrH$ is straightforward to resolve.
In the unsaturated atmospheres, the leading order behaviour of $\scrH$ is more complicated and the linearized equations gain a non-constant coefficient $\scrN$ that is rather sharp and requires significant resolution.
In either case,  one must provide an approximation to the Heaviside function that can be linearized.
The most common choices are $\tanh$ and $\erf$; both require a parameter $k$ that determines the steepness of the interface.
By comparing the NLBVP solutions to the analytic and asymptotic solutions, we can quantify the convergence of the approximate Heaviside function as a function of the artificial parameter $k$ and the physical parameter $\tau$, whose small size is an assumption of the Rainy Benard model.
Because $\scrH$ is not a function of spatial coordinates, it is possible to use a true piecewise function even when using spectral methods (D.\ Lecoanet, personal communication). However, this precludes the linear stability analysis we perform here.
We have tested both $\tanh$ and $\erf$ and find the latter to have modestly better properties, especially as related to spectral convergence; as reflected in equation~\ref{eq:smooth Heaviside} we use $\erf$ exclusively in the calculations below.

In the unsaturated cases, the problems posed by $\scrN$ are further exacerbated by the Legendre polynomials we use to discretize the $z$ direction, which cluster points preferentially away from the center of the interval, where the sharp discontinuity is located near $z_c$.
For unsaturated atmospheres, we thus use three matched domains, one from $0 < z_1 < z_c - z_\epsilon$, one from $z_c - z_\epsilon < z_2 < z_c$, and a third from $z_c < z_3 < 1$ where $z_\epsilon$ is chosen dynamically by the $\erf$ width parameter $k$. This allows the natural clustering of resolution for the Legendre polynomials to enhance the resolution in the region at which the solutions are changing most rapidly. 
Each domain has its own state variables, $\left\{p_i, \mathbf{u}_i, b_i, q_i\right\}$ for each domain $i \in \left\{1,2,3\right\}$. 
The standard boundary conditions at the top and bottom of the total domain, $z=0,1$, remain the same and are supplemented with a set of matching conditions at the interfaces: $p, b, q, \mathbf{u}$ and the first derivatives of $b, q, \mathbf{u}$ are all continuous.
The equations for all three layers are identical; as they simply promote numerical efficiency, in what follows we do not differentiate between them, referring only to $f_1$ for the perturbations to variable $f$.

In order to determine the onset of instability, we solve a series of linear eigenvalue problems for perturbations to the background atmospheres  described in section~\ref{sec:drizzle}. We first decompose all quantities into a static, $z$-dependent background and a time-dependent fluctuation, $f(x, z, t) = f_0(z) + f_1(x, z, t)$, assume a modal dependence in time and the horizontal direction
\begin{equation}
    f_1(x,t) = \hat{f}_1(z)\exp{(\omega t - i \vec{k}_\perp\cdot \vec{x}_\perp)}
    \label{eq:modal}
\end{equation}
with $\vec{x}_\perp = x\vec{\hat{e}}_x + y \vec{\hat{e}}_y$, and then solve equations for complex eigenvalue $\omega= \omega_r + i \omega_i$ and eigenfunctions $\hat{f}_1(z)$.
Inserting equation~\ref{eq:modal} into equations~\ref{eq:NS}--\ref{eq:continuity}, we keep terms up to $\mathcal{O}(f_1)$.
Most terms are straightforward, but a few require some care.  

The Clausius–Clapeyron relation becomes
\begin{equation}
q_s = \exp{(\alpha (T_0 + T_1))} \approx q_{s0} (1 + \alpha b_1)
\end{equation}
with
\begin{equation}
q_{s0} \equiv \exp{(\alpha T_0)}.
\end{equation}

Terms involving the Heaviside function have the form $A \scrH(A)$ which becomes
% \begin{align}
%     A\scrH(A) &= (A_0 + A_1)\scrH(A_0 + A_1) \nonumber \\
%     &\approx A_0\scrH(A_0) + A_1\left[\scrH(A_0) + \frac{1}{2} k^2 \sech^2(k A_0)\right] + \mathcal{O}(A_1^2)  
% \end{align}
\begin{align}
    A\scrH(A) &= (A_0 + A_1)\scrH(A_0 + A_1) \nonumber \\
    &= A_0\scrH(A_0) + A_1\left[\scrH(A_0) + 
       A_0\frac{k}{\sqrt{\pi}} \exp\Big(-k^2 A_0^2)\Big)\right] + \mathcal{O}(A_1^2).
       \label{eq:AHA}
\end{align}
A subtle aspect of the phase-change term is that in the vicinity of the transition, $A_0=(q0-q_{s0}) = \mathcal{O}(\epsilon) \lesssim \mathcal{O}(A_1)$, and owing to this the combined second term in the square brackets with amplitude $A_1 A_0$ is at order $\mathcal{O}(A_1^2)$ rather than order $\mathcal{O}(A_1)$.  As a consequence, it cannot be included without also consistently including other terms at $\mathcal{O}(A_1^2)$, which could likely be done with a careful asymptotic analysis.  We do not do so here, and instead only include terms that are formally $\mathcal{O}(A_1)$ in the linear equations
\begin{align}
    A\scrH(A) &= A_0\scrH(A_0) + A_1\scrH(A_0) + \mathcal{O}(A_1^2).
    \label{eq:AHA O(A1)}
\end{align}
The convergence of NLBVP solutions to analytic solutions in section~\ref{sec:drizzle} using this ordering suggests it is sufficiently accurate.

In total, the linear contribution of the phase-change nonlinearity to order $\mathcal{O}(A_1)$ becomes:
\begin{equation}
(q-q_s) \scrH(q-q_s) \approx \scrN(z) (q_1 - q_{s0} \alpha b_1)
\label{eq:linear phase}
\end{equation}
where the non-constant coefficient $\scrN(z)$ in equation~\ref{eq:linear phase} is
\begin{equation}
    \scrN(z) \equiv \scrH(q_0 - q_{s0}).
\end{equation}
The $z$ dependence of $\scrN(z)$ arises from $q_0(z)$ and $q_{s0}(z)$.

With the analytic solutions for the base state $b_0$ and $q_0$, the linear system is:
\begin{align}
    \grad \cdot \vec{u_1} &= 0 \label{eq:linear continuity} \\
    \frac{\partial \vec{u_1} }{\partial t} - \scrR \lap \vec{u_1} + \grad p_1 -  b_1 \vec{\hat{e}}_z  &= 0 \\    
    \frac{\partial b_1}{\partial t} - \scrP \lap b_1 + \vec{u} \cdot \grad b_0
    - \frac{\gamma}{\tau}(q_1 - q_{s0} \alpha b_1) \scrN(z) &= 0 \label{eq:linear buoyancy} \\  
   \frac{\partial q_1}{\partial t} -\scrS \lap q_1 + \vec{u} \cdot \grad q_0
   + \frac{1}{\tau}(q_1 - q_{s0} \alpha b_1) \scrN(z) &= 0 \label{eq:linear moisture}
\end{align}
We restrict ourselves to two-dimensional modes, $\vec{k}_\perp\cdot \vec{x}_\perp = k_x x$, $k_y = 0$.  

The critical Rayleigh number $\Rac$ and wavenumber $k_{x,c}$ are the values at which the maximum $\omega_r = 0$.
To find these values, we initially scan on a discrete $k_x$, $\Rayleigh$ grid.  We identify two solutions that bracket $\omega_{r, \mathrm{peak}}=0$ in $\Rayleigh$.  For each of these we use \verb+scipy.optimize.minimize+ to identify the peak growth rates for continuous $k_x$, which are the points of maximal $\omega_r$ below and above $\Rac$.  Constructing $\omega_{r,\mathrm{peak}} = F(\Rayleigh, k_x)$, we then interpolate $F$ to find an approximate $\Rac^\prime$, $k^\prime_{x,c}$ where $\omega_r = 0$.  This serves as the initial condition for another optimization sweep at $\Rayleigh=\Rac^\prime$ and continuous $k_x$, and the new maximum $\omega_r$ is used to update the brackets in $\Rayleigh$. We continue this process until $\omega_{r,\mathrm{peak}} = 0$ to within a specified tolerance.

\subsection{Saturated Atmospheres}
The stability of saturated atmospheres depends strongly on whether or not the background atmosphere is conditionally unstable or not.  In all atmospheres that we studied, there is exchange of stability, with all critical modes having $\omega_i = 0$.  
The spectrum for saturated atmospheres is quite similar to that of Rayleigh-Benard and thus rather uninteresting: it consists of a set of zero-frequency modes differing only in their growth or damping rates.  We do not find any oscillatory modes in these saturated atmospheres, and for this reason we do not plot their spectrum here.

\begin{figure}
    \centering
    \includegraphics[width=\linewidth]{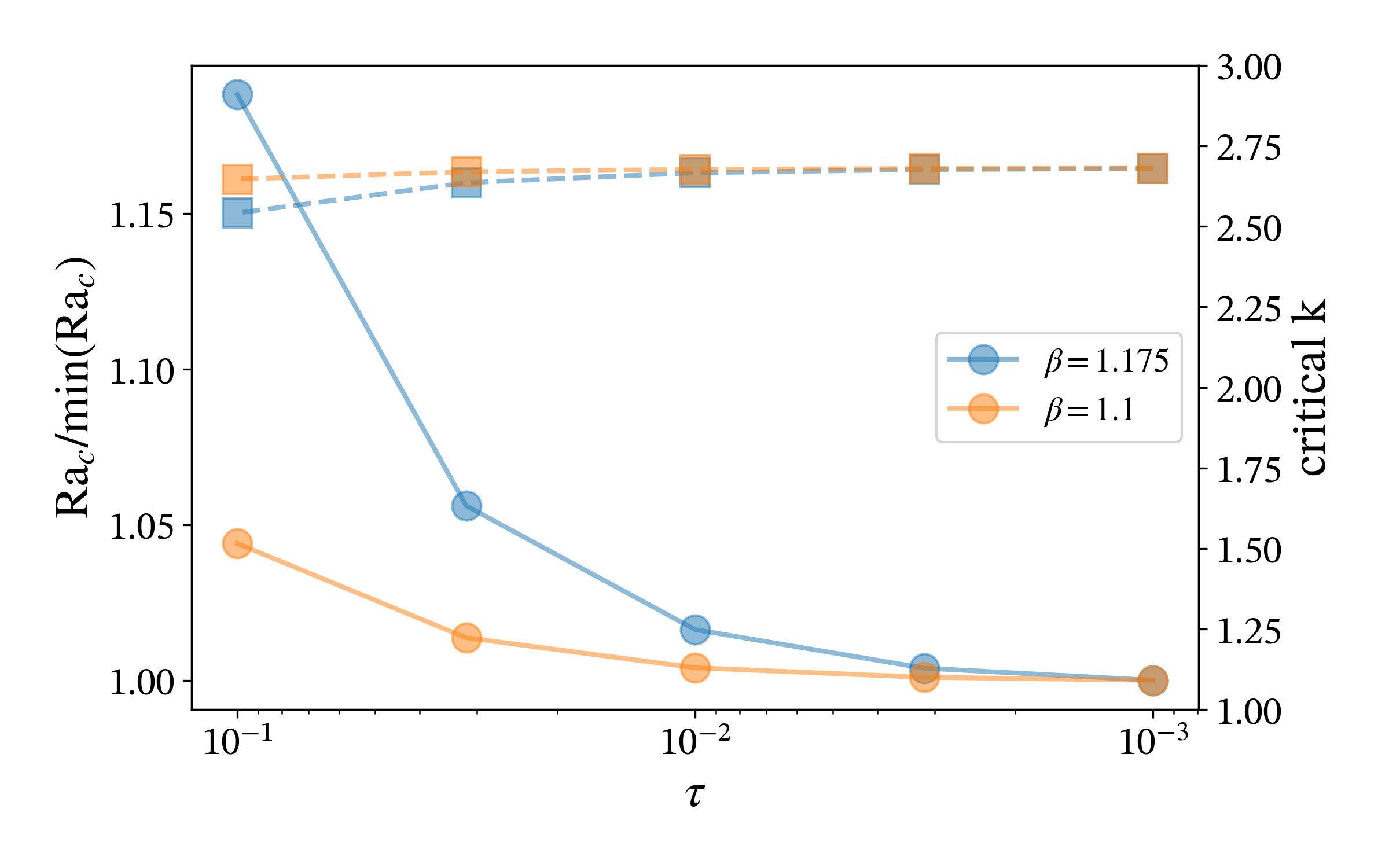}
    \caption{Relative critical $\Rayleigh_c$ numbers at $\gamma=0.19$, normalized by the smallest value at that $\beta$, (circles; left axis) and $k_c$ (squares; right axis) as functions of $\tau$ for saturated atmospheres.  Here we sample $\beta=1.175$ (blue, conditional instability) and $\beta=1.1$ (orange, unconditional instability), with $k=10^5$ in all cases.  As $\tau$ decreases, $\Rayleigh_c$ decreases and approaches a plateau value, as does $k_c$.}
    \label{fig:saturated_instability_tau}
\end{figure}

\begin{figure}
    \centering
    \includegraphics[width=\linewidth]{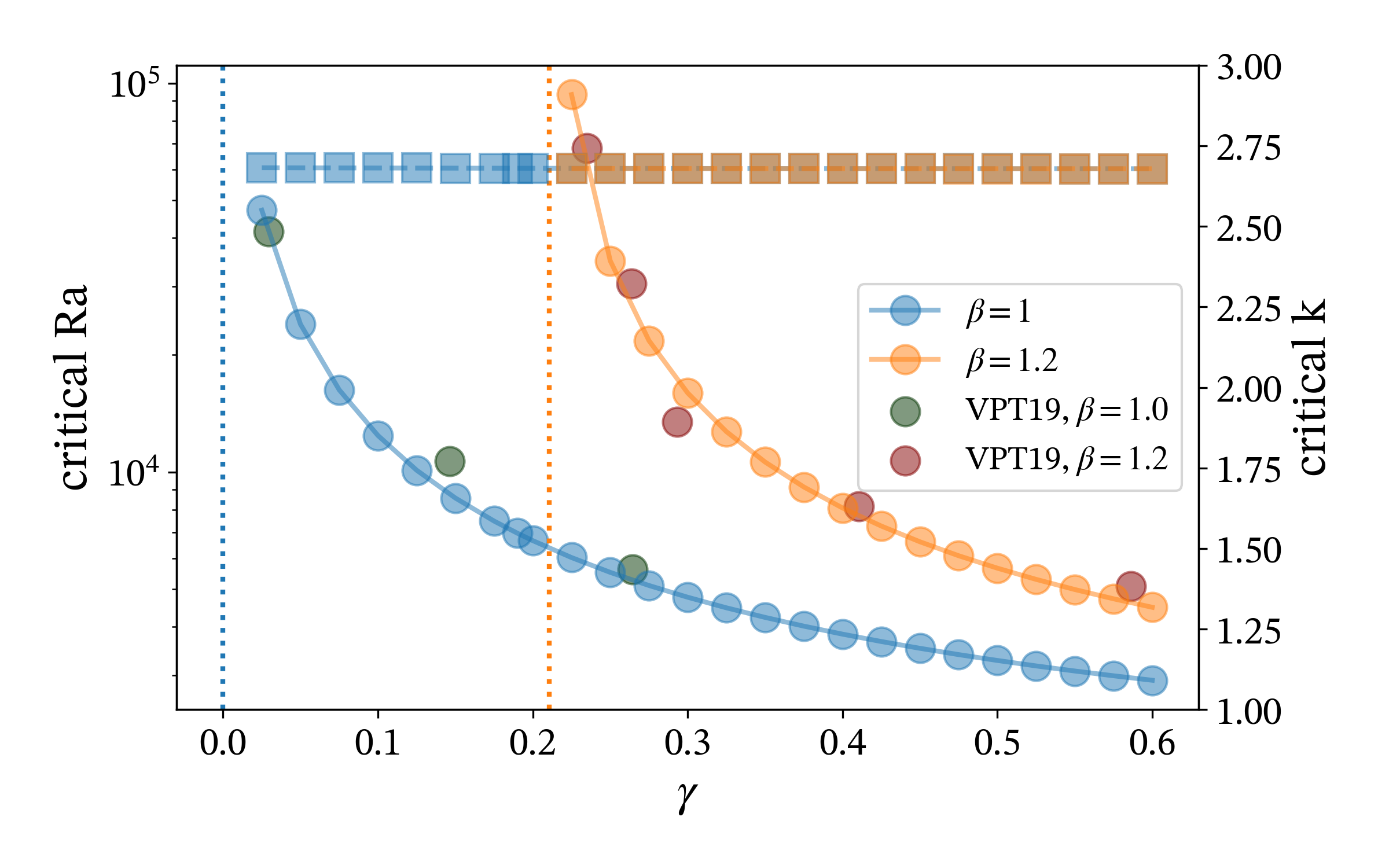}
    \includegraphics[width=\linewidth]{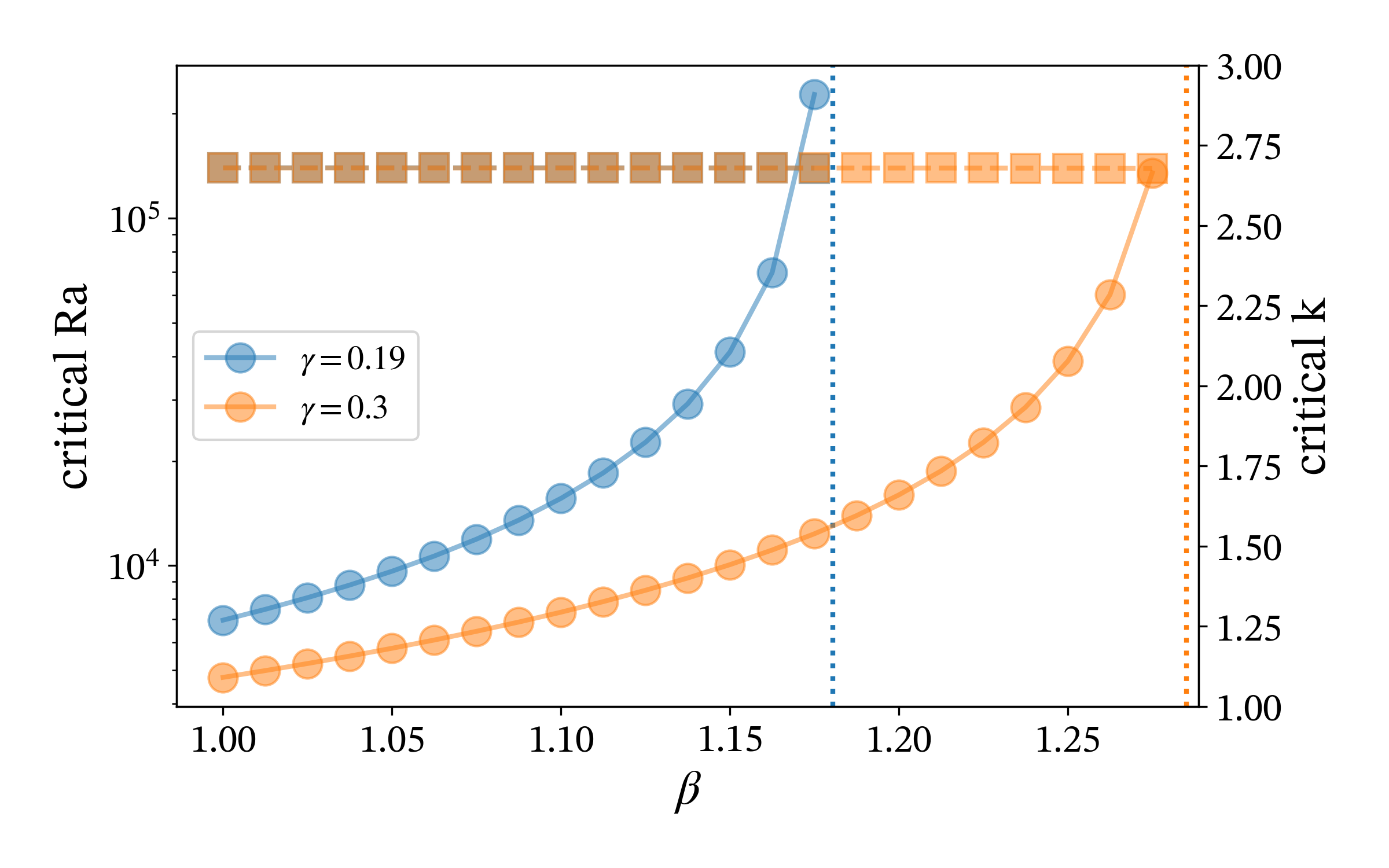}
    \caption{Critical Rayleigh numbers and $k_c$ for saturated atmospheres, with $\tau=10^{-3}$, $k=10^{5}$. The upper figure shows $\Rayleigh$ (circles; left axis) and $k_c$ (squares; right axis) as functions of $\gamma$ with $\beta=1.0$ (blue) and $\beta=1.2$ (orange); the lower figure shows the same quantities as a function of $\beta$ at $\gamma = 0.19$ (blue) and $\gamma = 0.3$ (orange). The upper figure also contains data from VPT19 as red and green circles. Note that these data have been scaled to account for a minor correction to the parameters in that paper (see appendix~\ref{sec:VPT19 correction}).}
    \label{fig:saturated_instability}
\end{figure}

Before we proceed, we first determine the dependence of critical $\Rayleigh$ on the condensation timescale $\tau$.  We test saturated atmospheres with $\beta=1.175$ (conditional instability) and $\beta=1.1$ (unconditional instability) and $\gamma=0.19$ for $\tau=[0.1,10^{-3}]$.  We set $k=10^5$ in all cases and $n_z=128$.  The results are shown in Figure~\ref{fig:saturated_instability_tau}.  The left axis shows that the critical $\Rac$ converges to less than $\sim 2\%$ at $\tau \simeq 10^{-2}$. When $\tau$ is larger, the critical $\Rayleigh$ is about 5--20\% larger than its converged value.  The critical wavenumber $k_c$ also depends on $\tau$; note that the right axis of figure~\ref{fig:saturated_instability_tau} shows \emph{absolute} values of $k_c$.  For the remainder of this work, we fix $\tau=10^{-3}$.  At this $\tau$ and $\gamma=0.19$, for the unconditionally unstable atmosphere at $\beta=1.1$, we find $\Rac \approx 1.56 \times 10^4$.  In the conditionally unstable atmosphere at $\beta = 1.175$ we find $\Rac \approx 2.27\times 10^5$.  In both cases the critical wavenumber $k_c \approx 2.68$.

\begin{figure}
    \centering
    \includegraphics[width=0.8\linewidth]{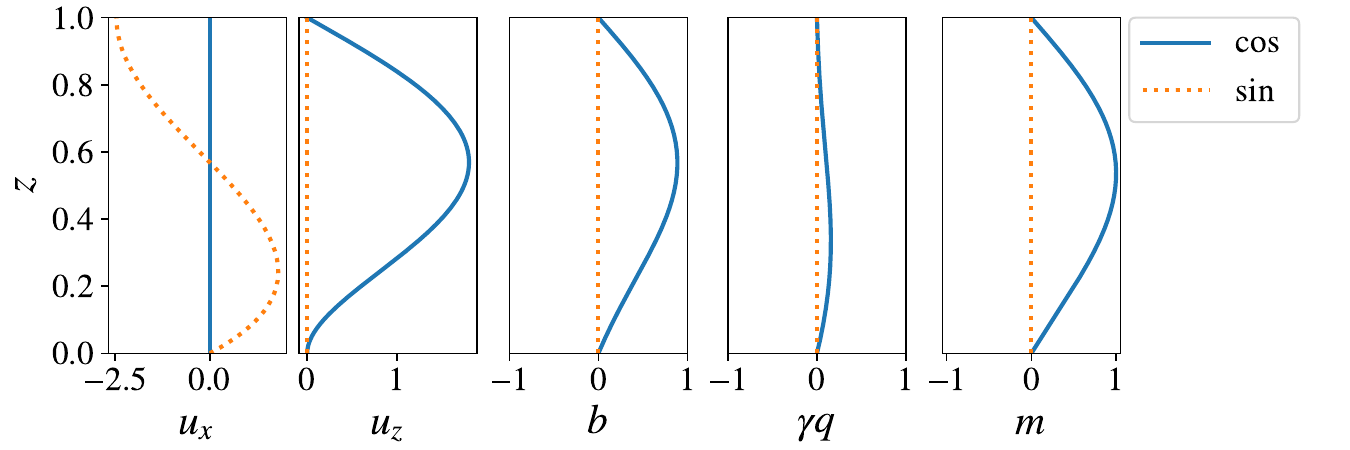}
        \includegraphics[width=0.8\linewidth]{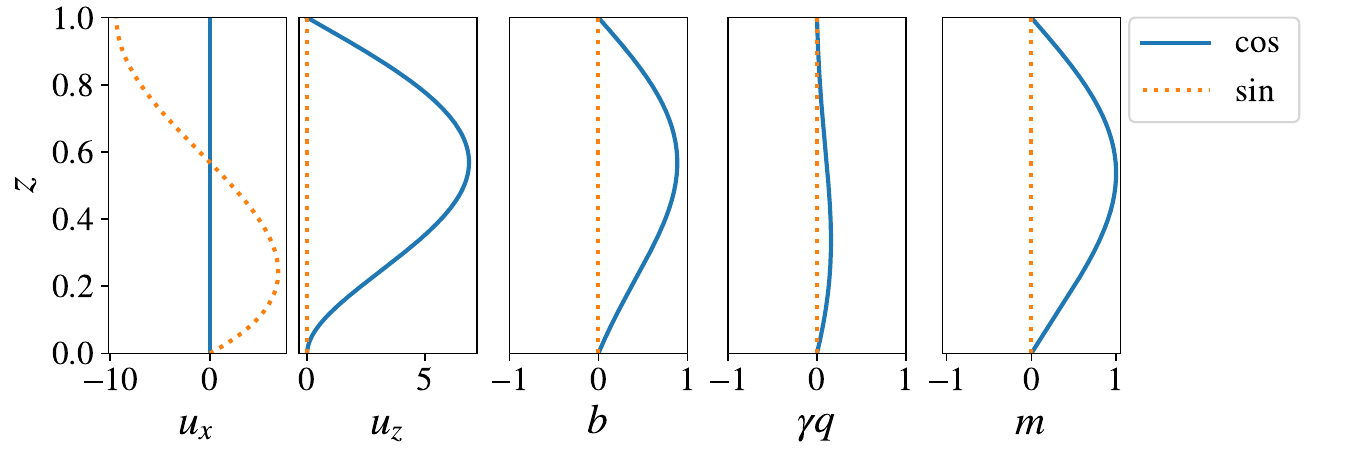}
    \caption{Eigenfunctions of fastest growing modes for saturated atmospheres at $\Rayleigh = \Rac$ top: unconditional instability ($\beta=1.1$), $\Rac \simeq 1.56 \times 10^4$, $k_c \simeq 2.68$, bottom: conditional instability ($\beta=1.175$), $\Rac \simeq 2.27 \times 10^5$, $k_c \simeq 2.68$. From left to right, perturbations to $u_x$, $u_z$, $b$, $\gamma q$, $m$.  The only difference between the two modes is that for $\beta=1.175$, the velocities are higher. All quantities are normalized such that $m = 1+ 0i$ at its maximum.}
    \label{fig:eigmode_sat}
\end{figure}

The dependence of $\Rac$ on $\gamma$ and on $\beta$ is given in Figure~\ref{fig:saturated_instability} where we show $\Rac$ for either two values of $\beta$ (top) or two values of $\gamma$ (bottom) to give a sense of the multi-dimensional parameter space.  The ideal stability limits are marked (dotted vertical lines), and $\Rac$ appears to diverge approaching these limits.  The critical wavenumber $k_c$ shows essentially no variation with either $\gamma$ or $\beta$.  In calculations at larger $\tau$ values (e.g., $\tau=0.1$, not shown), we have found that $k_c$ does vary with $\gamma$ and $\beta$, but that variation disappears as $\tau$ decreases below about $\tau\approx10^{-2}$ (e.g., Figure~\ref{fig:saturated_instability_tau}).

We include on the top panel of Figure~\ref{fig:saturated_instability} the results from VPT19, extracted with webplotdigitizer \citep{WebPlotDigitizer}.  We have scaled their results to account for a minor correction to the parameters reported in their paper (see appendix~\ref{sec:VPT19 correction} for details). 
The VPT19 results were found via timestepping nonlinear simulations and looking for exponential growth or decay, while our results come from the linear stability procedure described above.  The agreement is excellent at all $\beta$ and $\gamma$ studied by VPT19.

We plot the eigenfunctions at $\Rac$ for both $\beta = 1.1$ and $1.175$ in figure~\ref{fig:eigmode_sat}.  Generally, we find that $u_z$, $b$, $\gamma q$ and $m$ all share the same phase while $u_x$ is $\pi/2$ out of phase.  The $b$, $m$ and $u_z$ perturbations peak at similar heights while the $q$ perturbation peaks in the lower half of the domain.  The combined structure of $u_z$ and $u_x$ is very similar to the classic cellular patterns observed in linear Rayleigh-Benard convection.

Comparing the eigenfunctions at $\Rac$ for $\beta = 1.1$ and $1.175$, we see surprisingly little difference. The only visible difference between the two is the higher velocity perturbations $u_x$ and $u_z$ for the conditionally unstable case ($\beta=1.175$), in keeping with the much reduced diffusion at the significantly increased $\Rayleigh$. This suggests that while conditional instability dramatically stabilizes the system with respect to diffusive effects, the mechanism of instability remains the same: latent heat release causes an increase in buoyancy over the background.

\subsection{Unsaturated Atmospheres}
\begin{figure}
    \centering
    \includegraphics[width=0.8\textwidth]{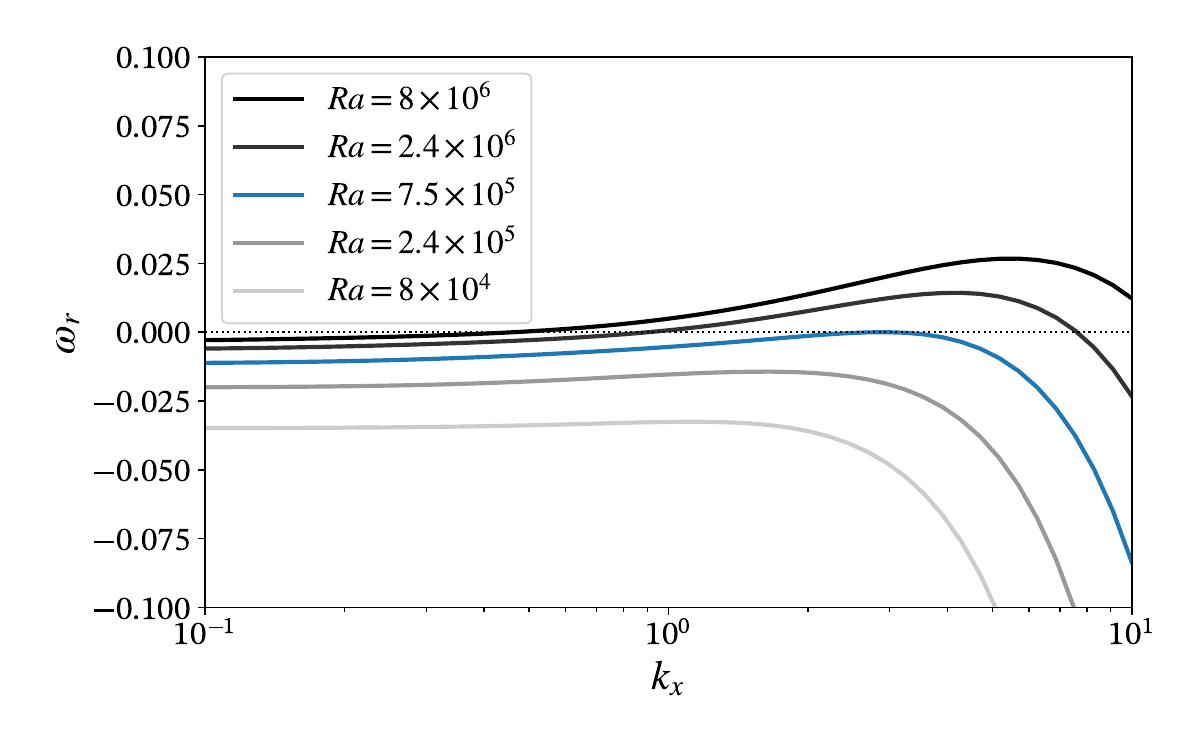}
    \caption{Growth rates as a function of wavenumber $k_x$ for five values of $\Rayleigh$ in a conditionally unstable, unsaturated atmosphere, with $\beta=1.1, \alpha=3, \gamma=0.19, q_0 = 0.6$. This background atmosphere is dry stable but moist unstable; the critical $\Rac \simeq 7.50\times10^5$ , with $k_c \simeq 2.89$, is shown in blue.}
    \label{fig:growth_vs_kx_beta_1.1_q0_0.6}
\end{figure}

\begin{figure}
    \centering
    \includegraphics[width=0.8\textwidth]{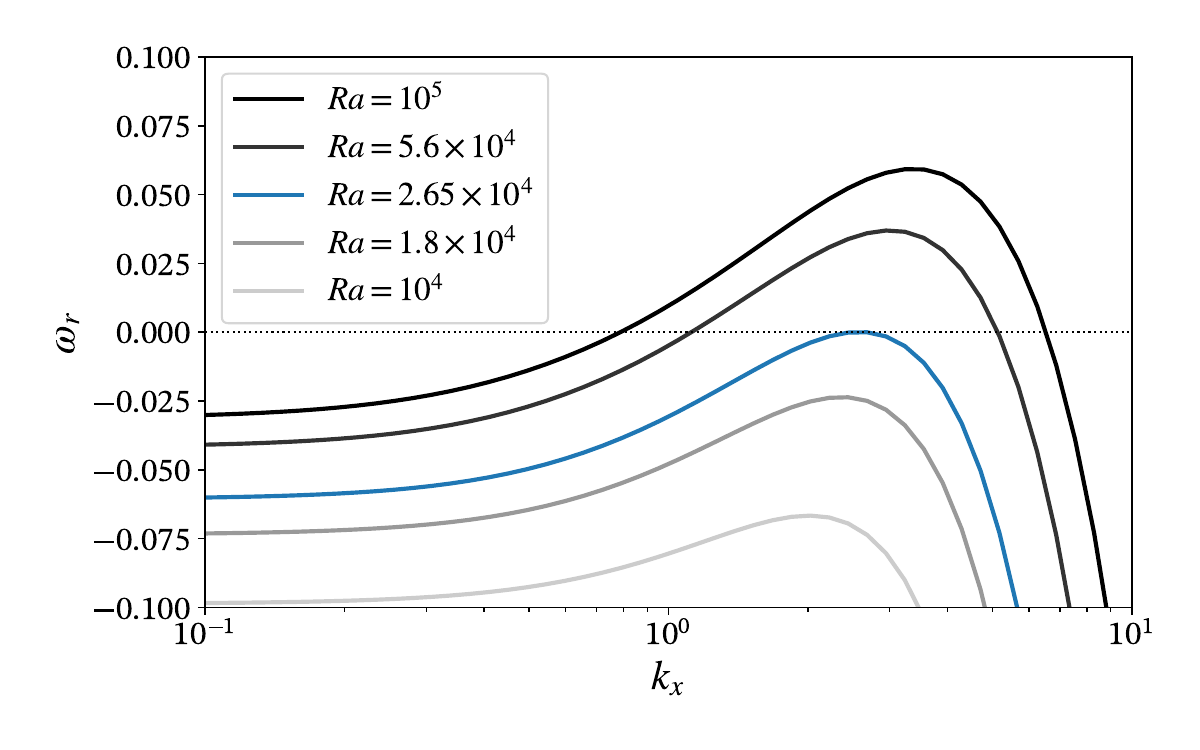}
    \caption{Growth rates as a function of wavenumber $k_x$ for four values of $\Rayleigh$ in an unconditionally unstable, unsaturated atmosphere, with $\beta = 1.05,\alpha=3, \gamma=0.19,q_0 = 0.6$. This background atmosphere is unstable to both dry and moist convection.  The critical $\Rac \simeq 2.65\times10^4$, with $k_c \simeq 2.58$, is shown in blue.}
    \label{fig:growth_vs_kx_beta_1.05_q0_0.6}
\end{figure}

We now turn to the unsaturated atmospheres with $\alpha = 3$, $\gamma = 0.19$, and $q_0 = 0.6$ (with $\tau=10^{-3}$ and $k=10^5$) and study their linear stability.
First, we determine the critical $\Rayleigh$ for a conditionally unstable atmosphere ($\beta = 1.1$) and an unconditionally unstable atmosphere ($\beta = 1.05$). The growth rates as a function of $k_x$ for these two cases are plotted in figures~\ref{fig:growth_vs_kx_beta_1.1_q0_0.6} and \ref{fig:growth_vs_kx_beta_1.05_q0_0.6} respectively.  Here, the solutions are more challenging to resolve, especially in the regions where the non-constant coefficient $\scrN(z)$ varies rapidly: even with the three matched domains for unsaturated atmospheres, we generally use higher resolutions (up to $n_z=512$ in total, with $n_{z,1}=n_{z,3}=128$ and $n_{z,2}=256$ in the matching layer).  These resolutions were sufficient for convergence here, but it is possible that more efficient calculations with lower resolutions could be completed.   

We start with the conditionally unstable atmosphere.  As can be seen in figure~\ref{fig:growth_vs_kx_beta_1.1_q0_0.6}, at $\Rayleigh>\Rac$ there is a range of $k_x$ where modes grow in amplitude ($\omega_r > 0$) while at $\Rayleigh<\Rac$ all modes are damped ($\omega_r < 0$).  The critical $\Rac \approx 723,000$ curve contacts $\omega_r=0$ at one point, the critical wavenumber $k_c$.  The fastest growing mode has $\omega_i=0$ and the instability proceeds via exchange of stability, as in the saturated atmosphere cases earlier.
The structure of these growth curves is broadly similar to those found in classical Rayleigh Benard convection.

We turn now to the unconditionally unstable atmosphere (figure~\ref{fig:growth_vs_kx_beta_1.05_q0_0.6}).  Here the behaviour is similiar, with growing modes when $\Rayleigh>\Rac$, decaying modes when $\Rayleigh<\Rac$, and instability at a critical $\Rac$ and wavenumber $k_c$ via exchange of stability.  Compared to the conditionally unstable atmospheres, here the growth curves are more peaked, including for cases with $\Rayleigh<\Rac$. As expected from the fact that $\beta = 1.05$ is unstable to both moist and dry convection, the critical $\Rac = 2.69\times10^4$ for this unconditionally unstable atmosphere is significantly lower in comparison to that for the $\beta = 1.1$ conditionally stable atmosphere, where it rises to $\Rac = 7.23\times10^5$. Curiously, however, the wavenumber at onset is hardly changed at all. We shall return to this point below.

\begin{figure}
    \centering
    \includegraphics[width=0.8\linewidth]{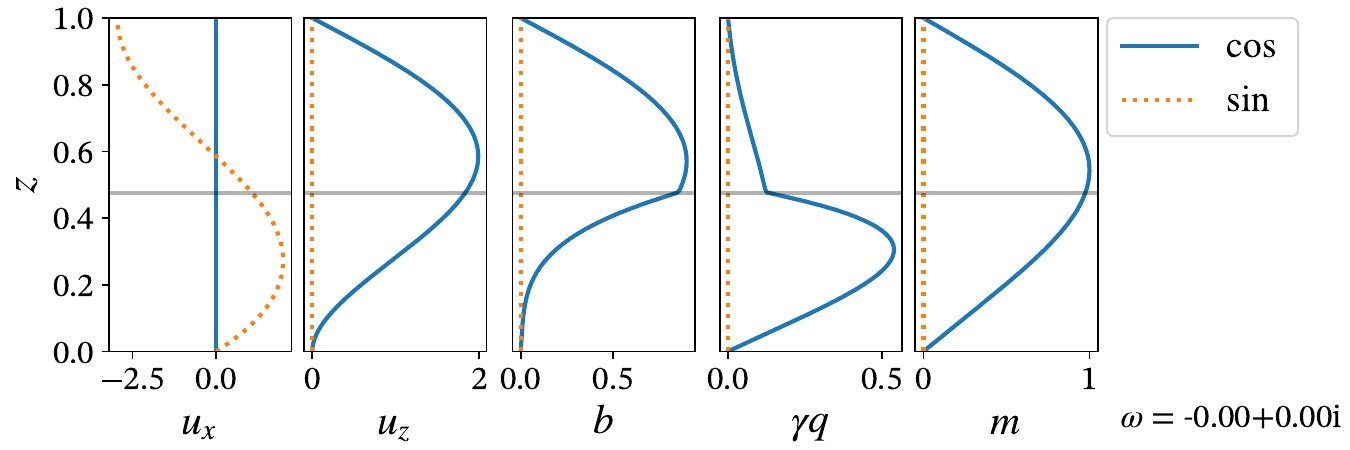}
    \includegraphics[width=0.8\linewidth]{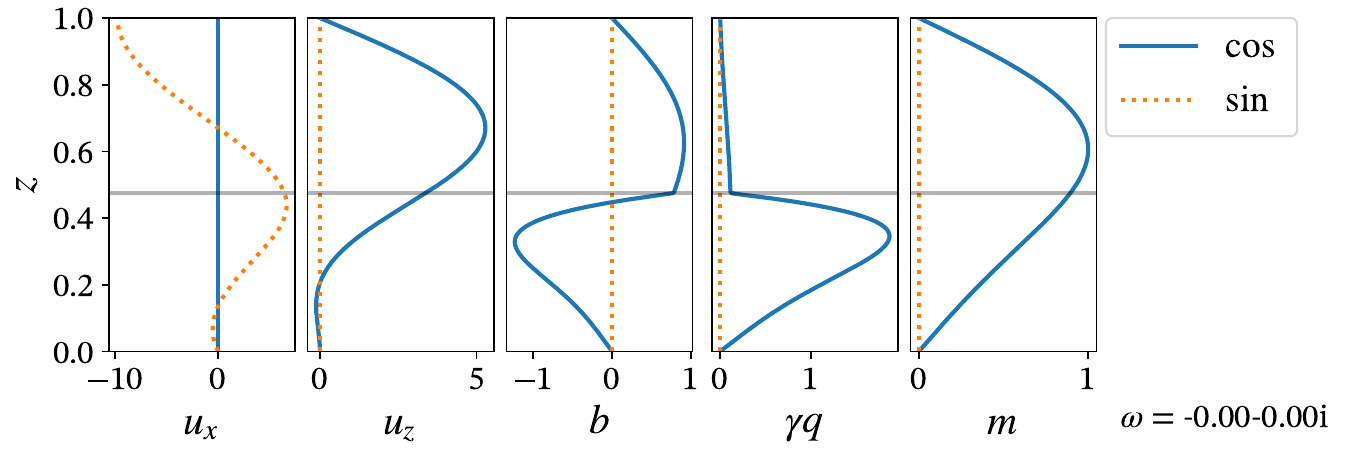}

    \caption{Eigenfunctions for fastest growing modes for unsaturated atmospheres at $\Rayleigh = \Rac$; top: unconditional instability ($\beta=1.05$), $\Rac \simeq 2.68 \times 10^4$, $k_c \simeq 2.57$ bottom: conditional instability ($\beta=1.1$), $\Rac \simeq 6.87 \times 10^5$, $k_c \simeq 2.60$. From left to right, perturbations to $u_x$, $u_z$, $b$, $\gamma q$, $m$.   All quantities are normalized such that $m = 1+ 0i$ at its maximum.}
    \label{fig:eigmode_unsat}
\end{figure}

We plot the eigenfunctions for the fastest growing mode at $\Rac$ for both $\beta = 1.05$ and $1.1$ in figure~\ref{fig:eigmode_unsat}.  The upper saturated portions of the atmospheres, above $z_c$, look very similar to our earlier results for fully saturated atmospheres, with $u_z$, $b$, $\gamma q$ and $m$ all share the same phase while $u_x$ is $\pi/2$ out of phase.  The most unstable modes span down into the lower, unsaturated part of the domain.  In that lower region the amplitude of $\gamma q$ grows substantially as here $q < q_s$ and no latent heat release modifies $b$.  The velocities $u_z$, $u_x$ and $m$ span the transition quite smoothly, while $b$ and $\gamma q$ show substantial structure at the transition $z=z_c$.  The combined structure of $u_z$ and $u_x$ is very similar to the classic cellular patterns observed in linear Rayleigh-Benard convection, though here that structure spans down into the unsaturated region.  In the conditionally unstable atmosphere, the most unstable mode has a small second cell located at the very lowest regions of the atmosphere ($z \lesssim 0.2$), and the $b$ fluctuations are generally of larger amplitude and different structure than the unconditionally unstable atmosphere.

\begin{figure}
    \centering
    \includegraphics[width=\linewidth]{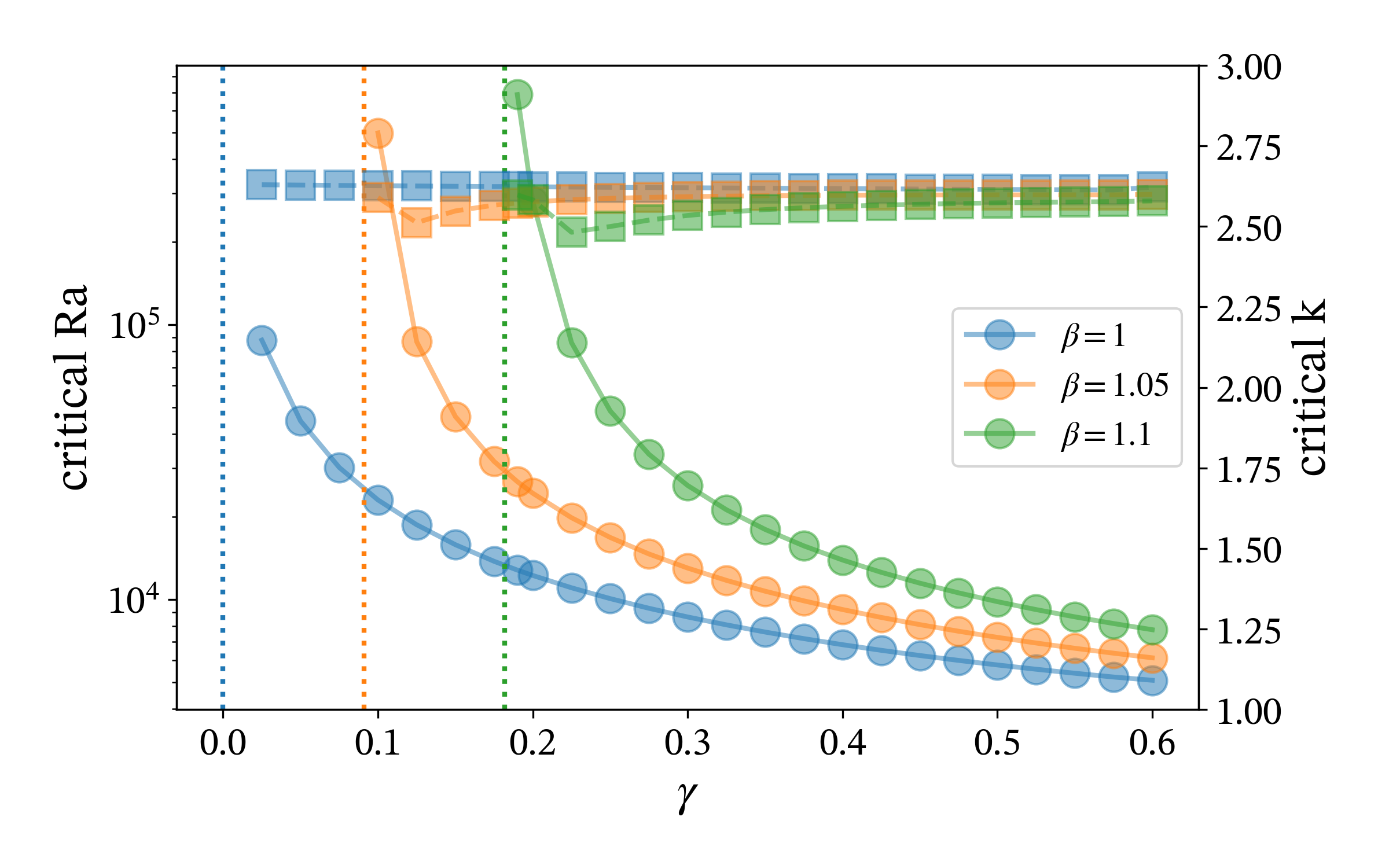}
    \includegraphics[width=\linewidth]{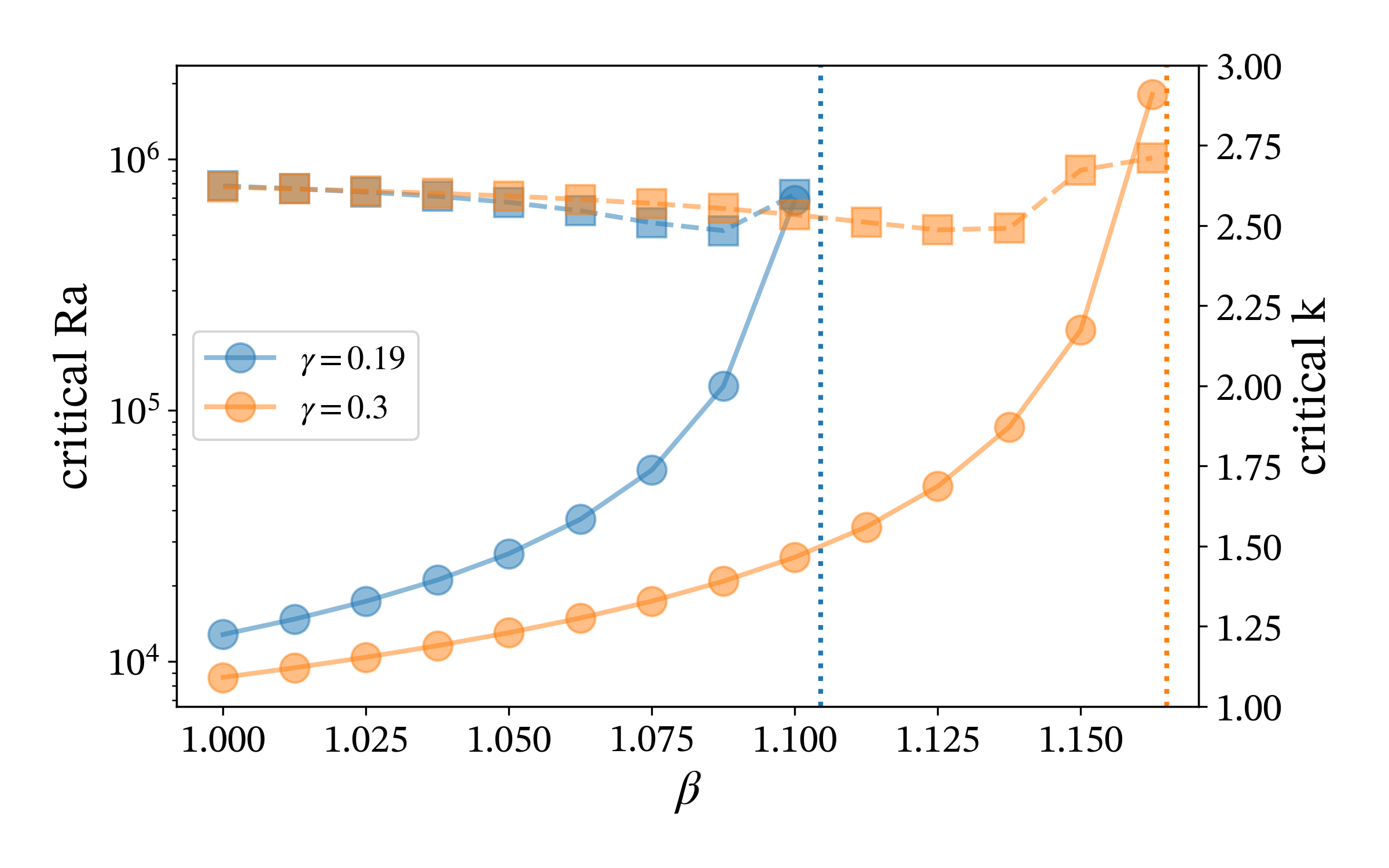}
    \caption{Critical Rayleigh numbers and $k_c$ for unsaturated atmospheres, with $\tau=10^{-3}$, $k=10^5$. The upper figure shows $\Rayleigh$ (circles; left axis) and $k_c$ (squares; right axis) as functions of $\gamma$ with $\beta=1.0$ (blue), $\beta=1.05$ (orange) and $\beta=1.1$ (green); the lower figure shows the same quantities as a function of $\beta$ at $\gamma = 0.19$ (blue) and $\gamma = 0.3$ (orange).}
    \label{fig:unsaturated_instability}
\end{figure}

The dependence of $\Rac$ on $\gamma$ and on $\beta$ for these unsaturated atmospheres is shown in Figure~\ref{fig:unsaturated_instability}.
Here, we include three values of $\beta$ (top) and the same two values of $\gamma$ (bottom) to illustrate the multi-dimensional parameter space.  Ideal stability limits are again marked (dotted vertical lines), and $\Rac$ appears to diverge approaching these limits.  The critical wavenumber $k_c$ shows much less variation, though there is more variation than was seen in the fully saturated atmospheres (Figure~\ref{fig:saturated_instability}).  Some ordering of $k_c$ with $\beta$ is visible in the top panel, with $\beta=1$ having a typical values of $k_c \approx 2.62$, $\beta=1.05$ having typical values of $k_c \approx 2.59$ and $\beta=1.1$ having typical values of $k_c \approx 2.57$. 
 As the critical value of $\gamma$ (top) or $\beta$ (bottom) is approached, $k_c$ appears to decrease and then increase.  These variations are generally small compared to the large variations in $\Rac$, and would likely diminish further at even smaller values of $\tau$.

\begin{figure}
    \centering
    \includegraphics[width=0.3\textwidth]{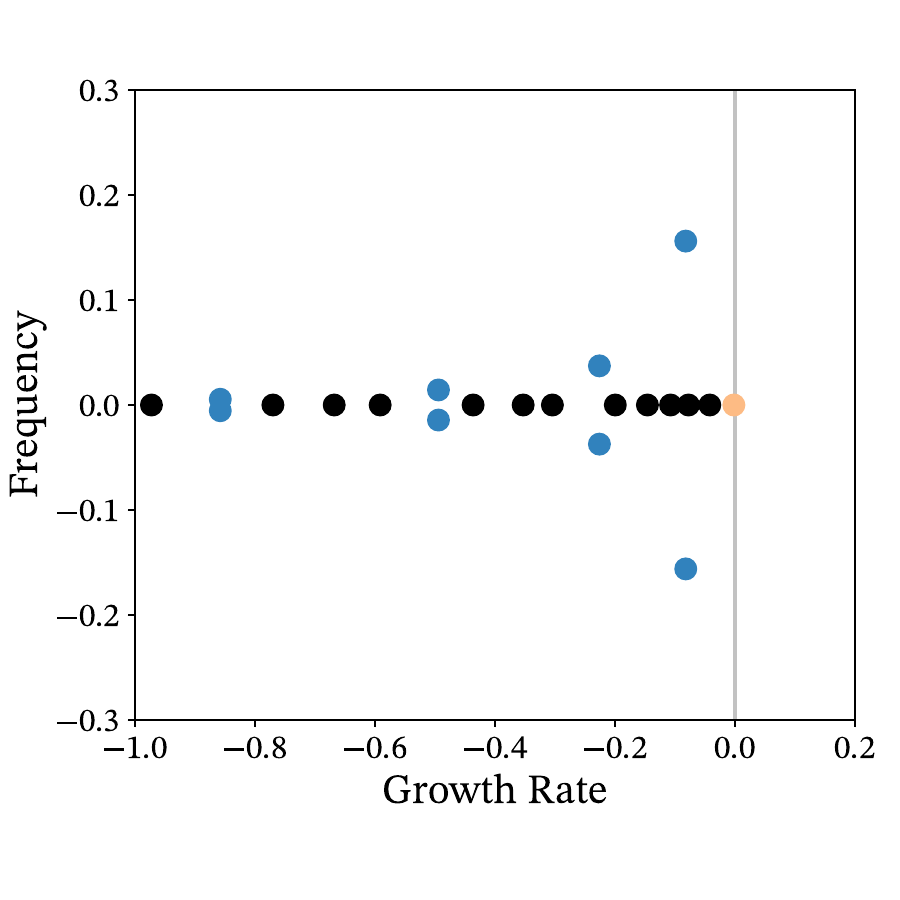}
    \includegraphics[width=0.3\textwidth]{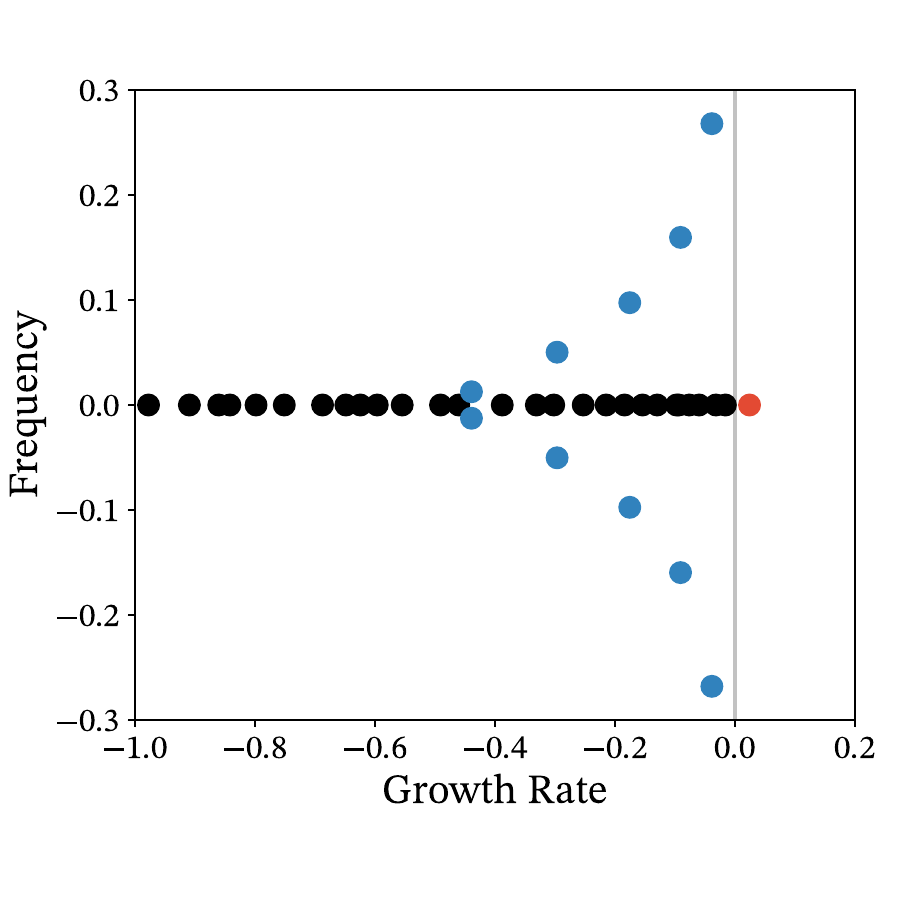}
    \includegraphics[width=0.3\textwidth]{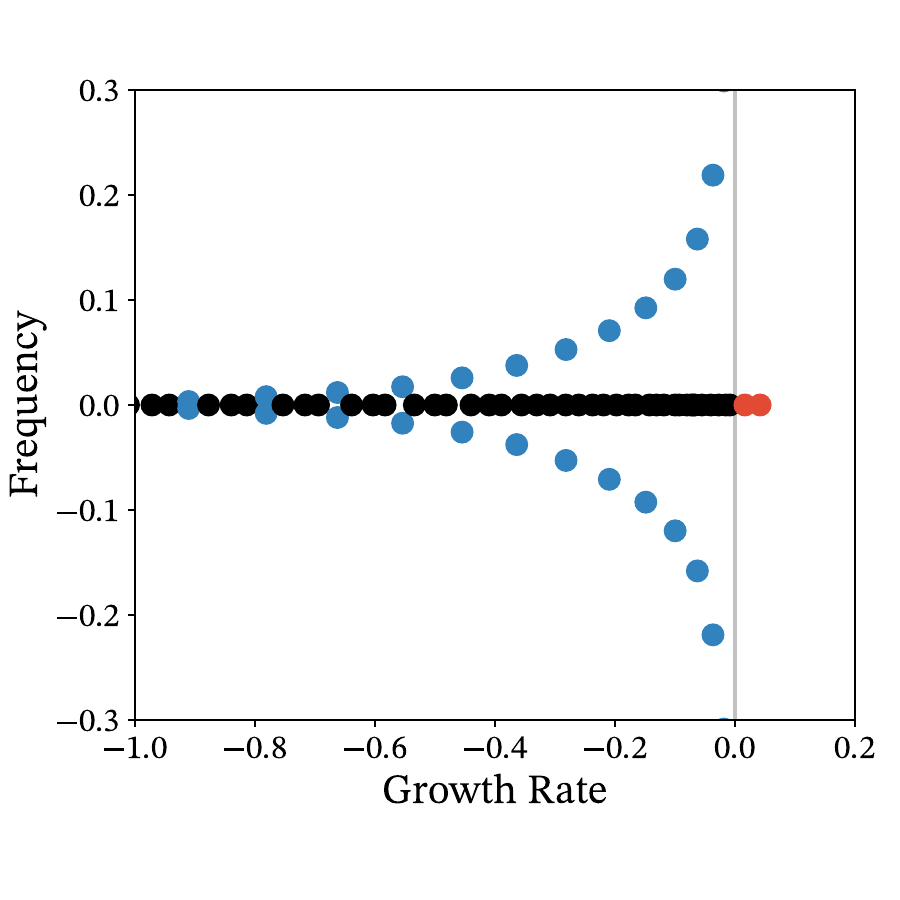}
    \caption{Eigenvalue spectrum $\omega = \omega_r + i \omega_i$ for $\Rayleigh = \Rac= 6.87\times10^5$, $k_x = 2.60$ (left), $\Rayleigh = 10 \Rac$, $k_x = 6.0$ (center), $\Rayleigh = 100 \Rac$, $k_x = 9.0$ (right) for an unsaturated atmosphere with conditional instability ($\gamma=0.19$, $\beta = 1.1$). $x$-axis shows the growth rate $\omega_r$; $y$-axis shows frequency $
    \omega_i$. Blue points indicate wave modes with non-zero frequencies; red points show growing modes; pale orange point shows the neutral mode.}
    \label{fig:unsat_beta1.1_spectrum}
\end{figure}

Next, we consider the eigenvalue spectrum of the unsaturated atmospheres, where the situation is more interesting than for the saturated cases.  
The spectrum for the conditionally unstable atmosphere with $\beta=1.1$ is shown in figure~\ref{fig:unsat_beta1.1_spectrum}.  
Shown are cases at (left panel), above (middle panel, $\Rayleigh = 10 \Rac$), and well above (right panel, $\Rayleigh = 100\Rac$) the critical $\Rac$.
Each case is computed at the wavenumber of the fastest growing mode.
In all cases, the onset of instability remains direct, with $\omega_i=0$ for the fastest growing mode.

At the highest $\Rayleigh=100\Rac$, two different growing modes are visible above onset.  The fastest growing mode has a single cell in the upper, saturated portion of the atmosphere, as in Figure~\ref{fig:eigmode_unsat}, while the slower growing mode is the next overtone, with two cells in $z$ in the saturated region.
In all three cases, new branches of damped oscillatory modes with $\omega_i \neq 0$ appear in the system.
The waves are not destabilized in the region of parameter space above onset: though the oscillatory branches become less damped, they also move to higher frequencies and do not seem likely to cross the $\omega_r = 0$ axis.

\begin{figure}
    \centering
        \includegraphics[width=0.3\textwidth]{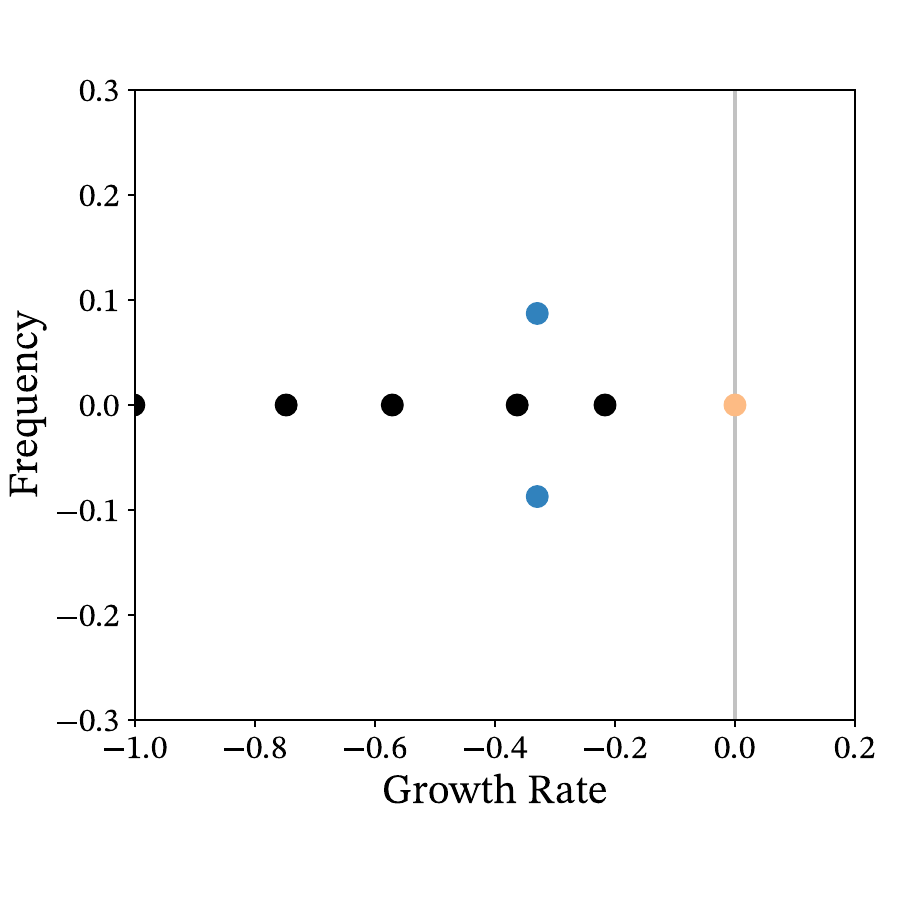}
        \includegraphics[width=0.3\textwidth]{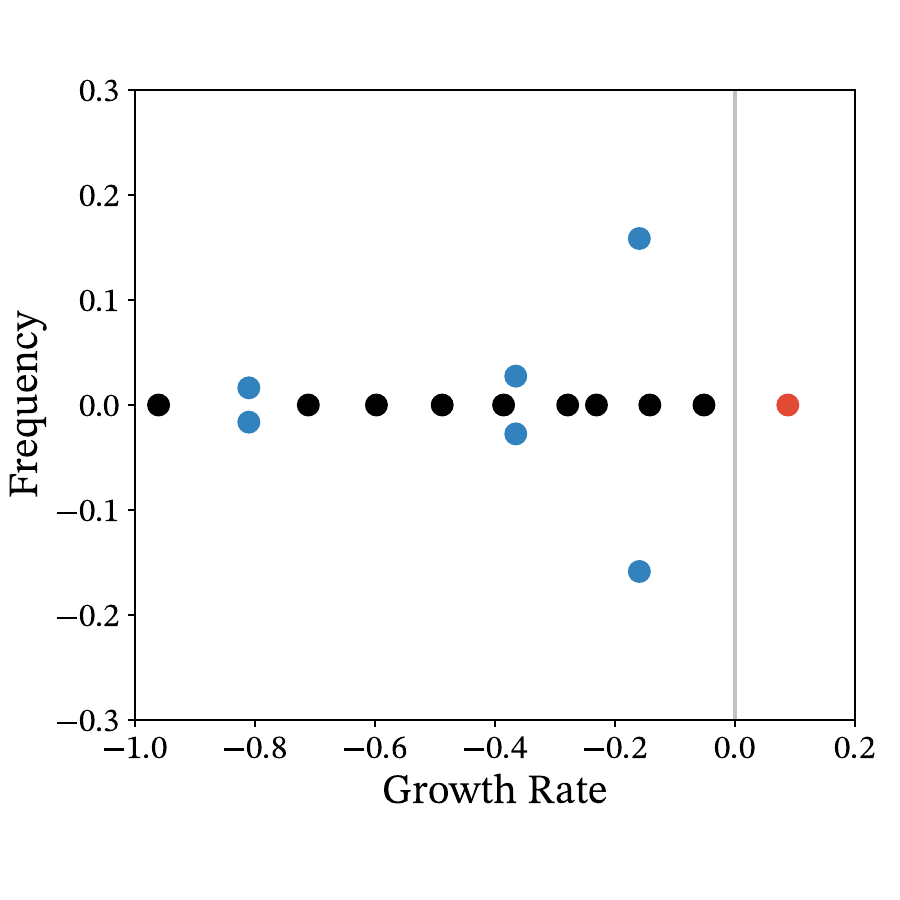}
        \includegraphics[width=0.3\textwidth]{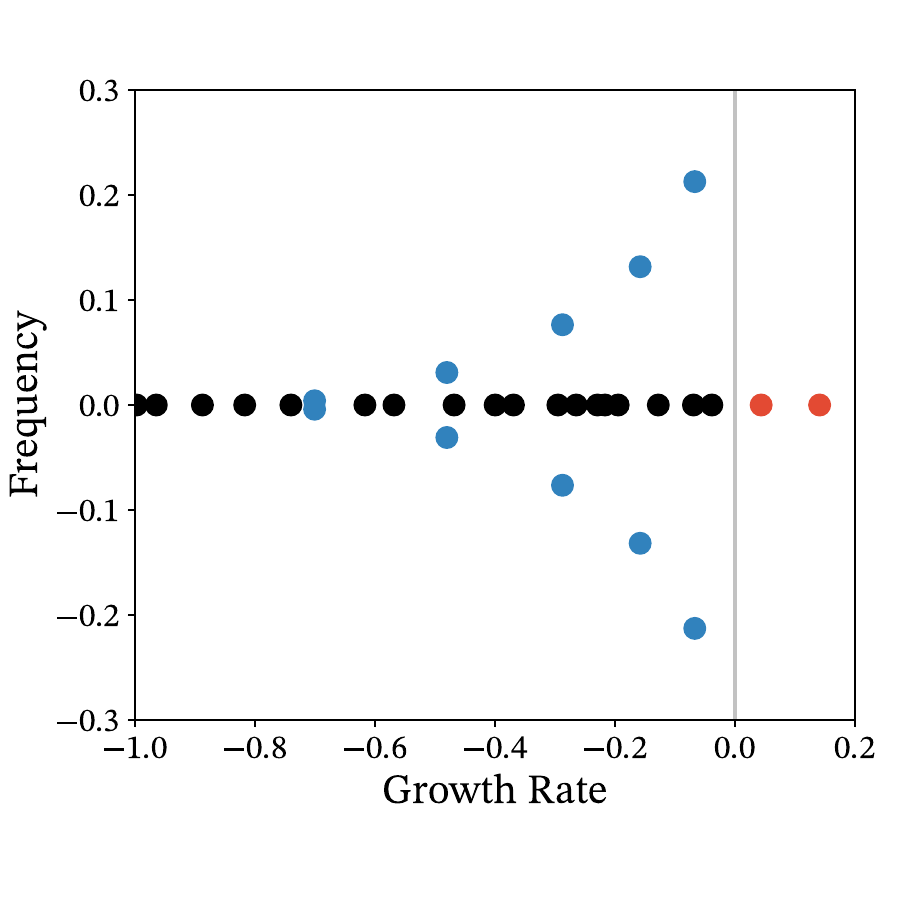}
    \caption{Eigenvalue spectrum $\omega = \omega_r + i \omega_i$ for $\Rayleigh = \Rac= 2.68\times10^4$, $k_x = 2.57$ (left), $\Rayleigh = 10 \Rac$, $k_x = 4.0$ (middle) and $\Rayleigh = 100 \Rac$, $k_x = 6.5$ (right) for an unsaturated atmosphere with unconditional instability ($\gamma=0.19$, $\beta = 1.05$). $x$-axis shows the growth rate $\omega_r$; $y$-axis shows frequency $
    \omega_i$. Each spectrum is at the $k_x$ corresponding to the peak growth rate at that $\Rayleigh$. Blue points indicate wave modes with non-zero frequencies; red points show growing modes; pale orange point shows the neutral mode.}
    \label{fig:unsat_beta1.05_spectrum}
\end{figure}

The spectrum for the unconditionally unstable atmosphere with $\beta=1.05$ is shown in figure~\ref{fig:unsat_beta1.05_spectrum}.  
Shown are cases at (left panel), above (center panel, $\Rayleigh = 10 \Rac$), and well above (right panel, $\Rayleigh = 100 \Rac$) the critical $\Rac$.
Each spectrum is computed at the $k_x$ of the fastest growing mode.
Here again all cases have direct onset of instability, with $\omega_i=0$ for the fastest growing mode.
At the highest $\Rayleigh=100\Rac$, two different growing modes are visible above onset, with larger separation than in the $\beta=1.1$ atmosphere.
Oscillatory branches are visible in all cases, with more such modes at higher $\Rayleigh$.

\begin{figure}
  \centering
\includegraphics[width=0.45\textwidth]{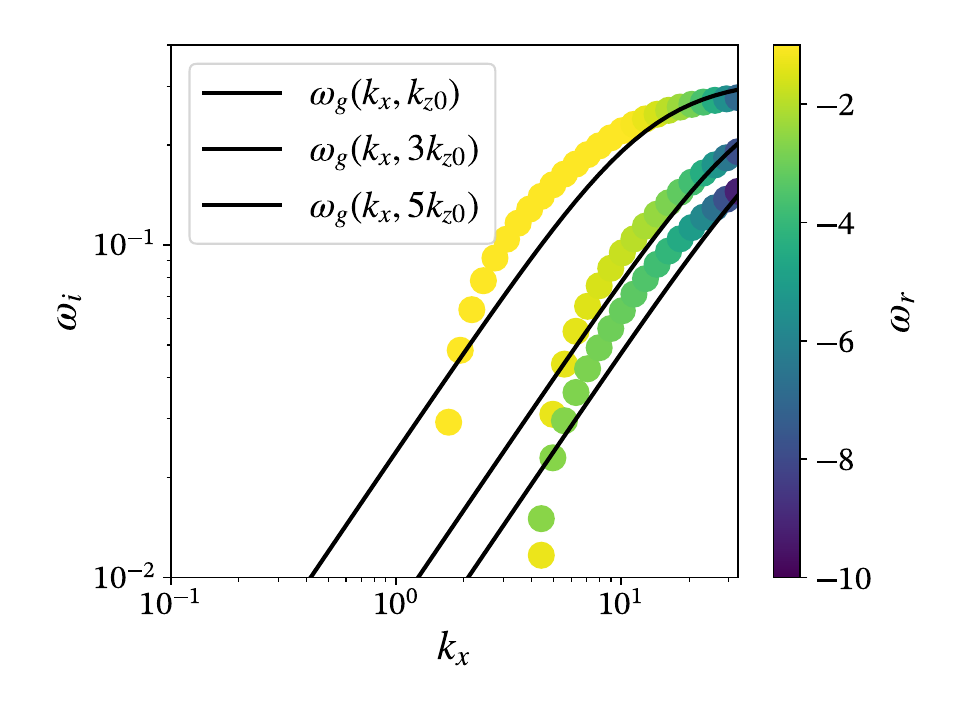}
\includegraphics[width=0.45\textwidth]{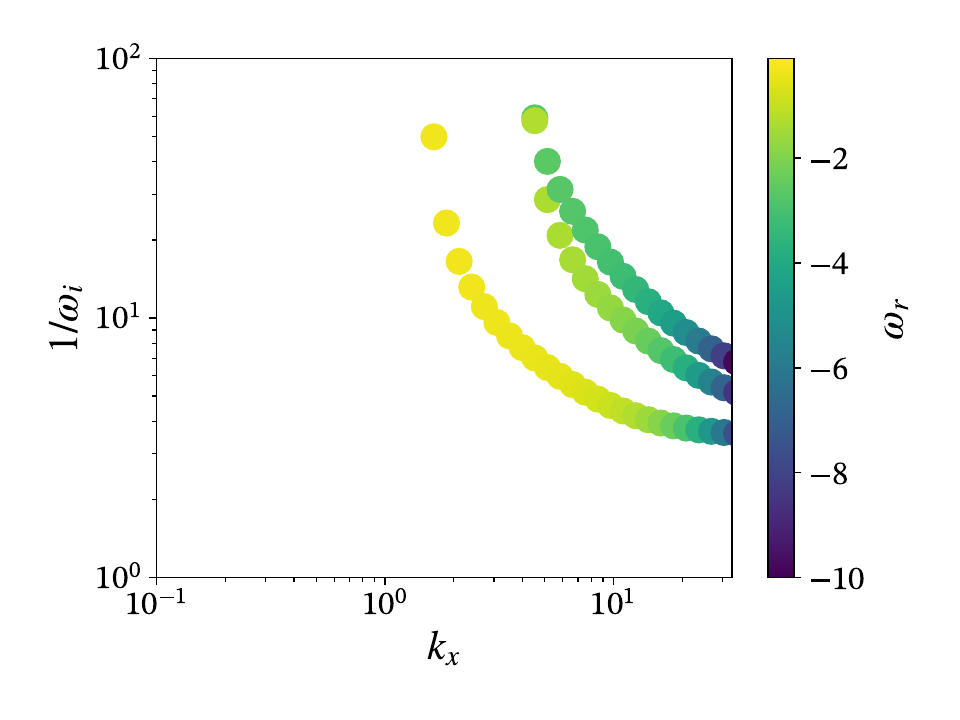}
  \caption{Wave frequency $\omega_i$ vs $k_x$ (left) and period (right) for an unconditionally unstable atmosphere ($\beta = 1.05$) at $\Rayleigh=\Rac \approx 2.65\times10^4$. The color of each point gives its damping rate $\omega_r$. The solid lines in the left panel shows the analytic dispersion relation for dry internal gravity waves $\omega_g(k_x, k_z) = N_b k_x/\sqrt{k_x^2 + k_z^2}$ for $k_{z0} = 2\pi/z_c$, $3k_{z0}$, and $5k_{z0}$. }
  \label{fig:omega_vs_k_beta1.05}
\end{figure}

In order to understand the nature of these waves, we plot their frequency as a function of horizontal wavenumber $k_x$ in figure~\ref{fig:omega_vs_k_beta1.05} for the conditionally unstable case $\beta = 1.05$ and at $\Rayleigh = \Rac \approx 2.8\times 10^{4}$.  Shown at left is a frequency diagram, while at right is a period diagram. 
 Several branches of oscillatory modes are visible, and all are damped ($\omega_r < 0$).
 At high $k_x$, and for modes with periods above the lowest period, the spacing of modes at fixed $k_x$ appears to be approximately constant in period.  This is a well-known characteristic of internal gravity waves \citep[e.g.][eq. 2.2.7]{Turner_1973}.
The black line shows the dispersion relation for Boussinesq internal gravity waves for $N_b = \partial_z b_0 \simeq 0.1$ (as seen in figure~\ref{fig:unsaturated_atmospheres}) and $k_z = 2\pi/z_c$. The gravity wave dispersion relation provides a reasonable approximation to the numerically computed frequencies when using the size of the \emph{unsaturated} region, strengthening our interpretation as moisture modified gravity waves.
The branches of modes at successively lower $\omega_i$ (at fixed $k_x$) represent modes with increasing structure in the $z$ direction; it is interesting to note that they appear best fit by odd multiples of $k_{z0}$.

\begin{figure}
    \centering
    \includegraphics[width=0.8\linewidth]{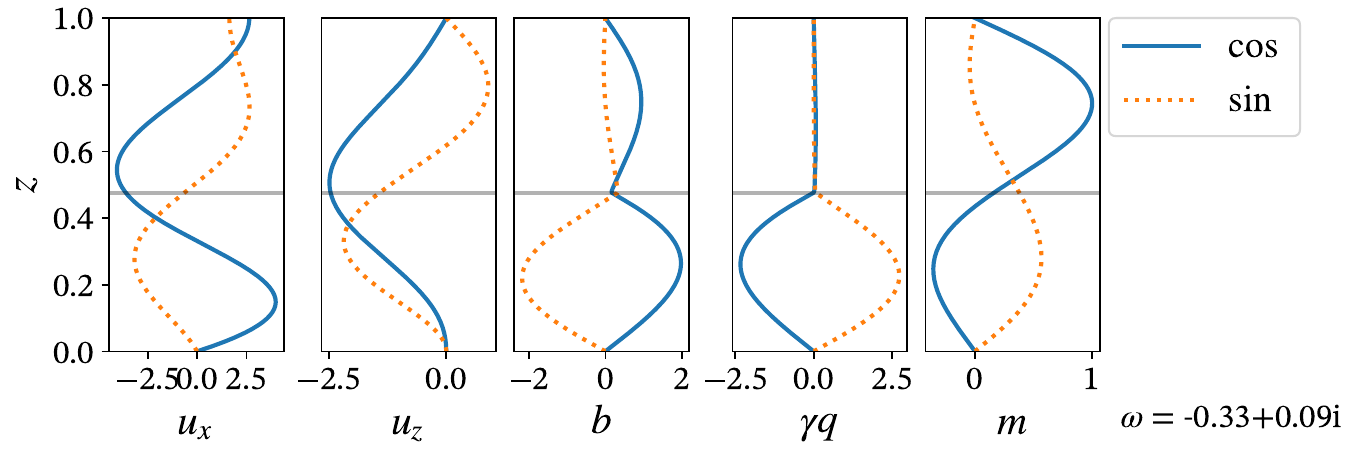}
    \includegraphics[width=0.8\linewidth]{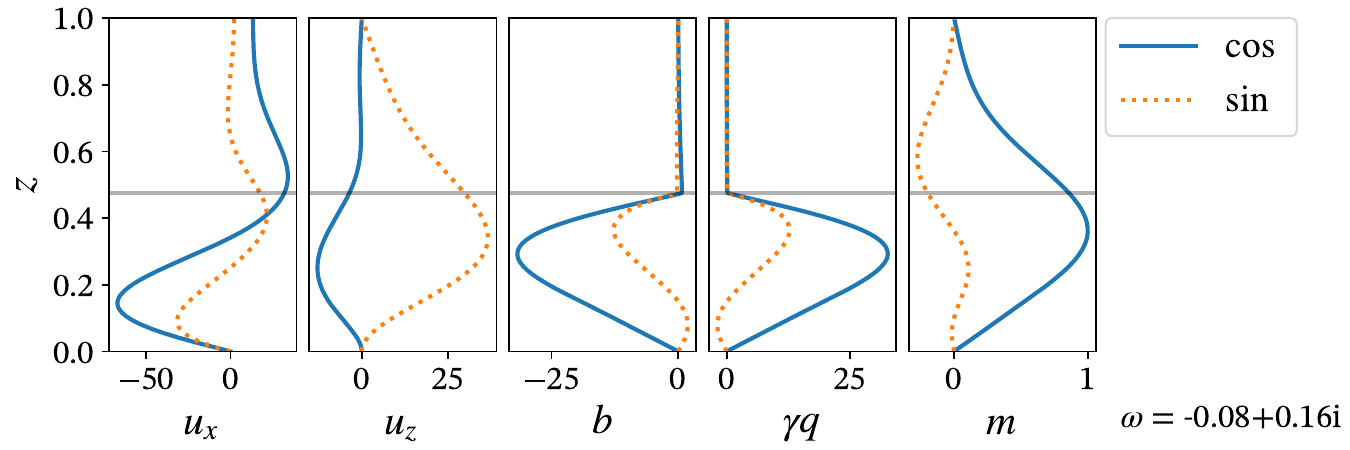}
    \caption{Eigenfunctions for highest frequency waves in the same atmospheres as in figure~\ref{fig:eigmode_unsat} at $\Rayleigh = \Rac$; top: unconditional instability ($\beta=1.05$), bottom: conditional instability ($\beta=1.1$). }
    \label{fig:eigmode_wave_unsat}
\end{figure}

The eigenfunctions for the highest frequency oscillatory modes in the unconditionally unstable atmosphere ($\beta=1.05$) and in the conditionally unstable atmosphere ($\beta = 1.1$) are shown in figure~\ref{fig:eigmode_wave_unsat}. 
Here the modes have more complex phase relationships than were present for the fastest growing modes (figure~\ref{fig:eigmode_unsat}), but some patterns hold:
the velocities $u_x$, $u_z$, and the moist static energy $m$ smoothly span between the saturated region above $z_c$ and the unsaturated region below, while the buoyancy $b$ and humidity $q$ show more structure in the vicinity of the transition at $z_c$.
It is striking that the profiles of $m = b + \gamma q$ are clearly continuous in figure~\ref{fig:eigmode_wave_unsat} across the $z=z_c$ transition, while the profiles of $b$ and $\gamma q$ appear at first to be discontinuous.  This is especially visible in the conditionally unstable atmosphere, where $b$ and $\gamma q$ appear to be zero in the saturated region of the atmosphere.  
However, there is no inconsistency.  Both $b$ and $\gamma q$ retain amplitude in the saturated region above $z=z_c$ due to incomplete cancellation, and the plotting is visually dominated by their very large amplitudes in the lower, unsaturated region of the atmosphere.

It is interesting that no such oscillatory modes are found in the fully saturated atmospheres, but the structure of the eigenfunctions in the unsaturated case sheds some light on that.
The eigenfunctions of the waves are always such that the buoyancy and humidity perturbations are oppositely signed.
This means that where positive buoyancy perturbations $b$ occur, the atmosphere is drier than average and latent heat is not being added. However, where the buoyancy perturbation is \emph{negative}, there is a increase in $q$ and the atmosphere is more moist.  This means that there is latent heat being added to the system, counteracting the lower buoyancy. In the lower unsaturated region of the atmosphere below $z_c$, $q < q_s$ and the moisture does not condense out.  As such, $q$ is decoupled from the buoyancy, rendering it unable to damp the wave. However, in the saturated portion of the atmosphere above $z_c$ the Heaviside function is essentially always on, and so the positive humidity perturbation couples directly to the buoyancy, adding additional buoyancy in the portion where the negative $b$ would otherwise provide a restoring force.  This acts to damp the waves in that region.  

\begin{figure}
    \centering
    \includegraphics[width=0.45\linewidth]{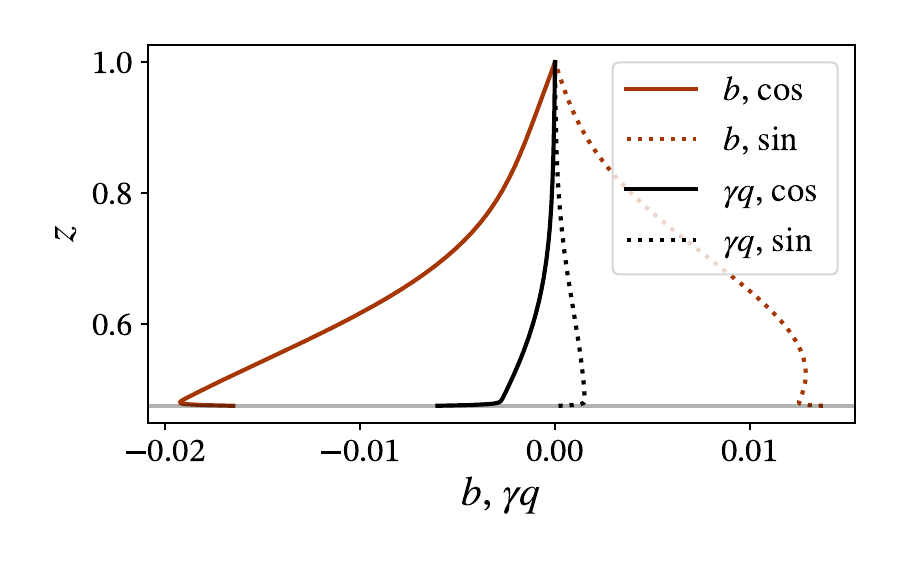}
    \includegraphics[width=0.45\linewidth]{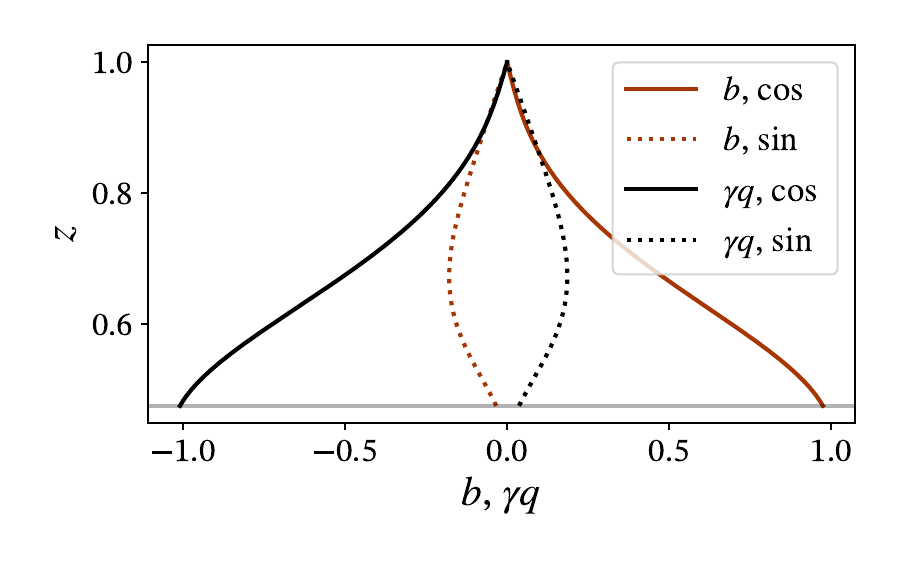}

    \caption{Here we show an experiment to verify the proposed latent heat wave-damping mechanism.  Zoom in on eigenfunctions for $b$ (orange red) and $\gamma q$ (black) in the saturated regions of partially-unsaturated atmosphere ($z > z_c$) for highest frequency waves shown in Figure~\ref{fig:eigmode_wave_unsat} at $\Rayleigh = \Rac$ for the conditionally unstably atmosphere ($\beta=1.1$). The critical level $z_c$ is indicated in grey, and here the modes are normalized by the max amplitude of $b$ in the domain, rather than $m$.  At left are the eigenfunctions for the wave equations with $\gamma=0.19$ and normal full coupling between $b$ and $q$.  The amplitudes here are very small, indicating a high degree of suppression of both $b$ and $q$.  At right are the eigenfunctions for $b$ and $\gamma q$ for wave equations where there is no coupling between $b$ and $q$ by setting $\scrN = 0$, corresponding to taking $\tau \rightarrow \infty$.  Here the $q$ fluctuations are almost exactly out of phase with the $b$ fluctuations, and the $b$ amplitudes remain large.}
    \label{fig:eigmode_wave_unsat_b_and_q_zoom}
\end{figure}

In these unsaturated atmospheres, the unsaturated region of the atmosphere below $z_c$ gives the waves a region to continue propagating without this damping.  In the fully saturated atmospheres studied earlier, there is no such dry region, and the waves are everywhere damped.  This explains why no linear oscillatory modes are found in the fully saturated atmospheres.

We tested this hypothesis by running a set of calculations in a conditionally unstable ($\beta = 1.175$ where $\nabla b_0 > 0$ everywhere), fully saturated atmosphere with the background state computed as normal at $\gamma = 0.19$ but with $\tau \rightarrow \infty$ by setting $\scrN(z) = 0$.  This decouples $b$ and $q$ and renders $q$ a passive scalar with sign opposite that of $b$ due to their respective oppositely-signed background gradients (see Figure~\ref{fig:saturated_atmospheres}).  While this is not self-consistent, it removes the latent heat and condensation terms from the dynamical equations.  Spectra of these modified equations in fully saturated atmospheres show oscillatory modes, in contrast to solutions with coupling between $q$ and $b$.  This suggests that the gravity waves are present in fully saturated atmospheres, but are damped by the negative feedback from the latent heat coupling of $q$ and $b$.  

To further test this hypothesis, we conducted the same experiment in the conditionally unstable, partially-unsaturated atmosphere used in Figure~\ref{fig:eigmode_wave_unsat} ($\gamma=0.19$, $\beta=1.1$).  In Figure~\ref{fig:eigmode_wave_unsat_b_and_q_zoom}, we zoom in on the eigenfunctions for $b$ and $q$ in the saturated part of that atmosphere at $z>z_c$.  The left panel shows the normal, full solution, while the right panel shows the solution with $\scrN(z)=0$.  The eigenfunctions in Figure~\ref{fig:eigmode_wave_unsat_b_and_q_zoom} are normalized differently than those in Figure~\ref{fig:eigmode_wave_unsat}, here using the maximum amplitude of $b$ rather than $m$.  In the left panel of Figure~\ref{fig:eigmode_wave_unsat_b_and_q_zoom}, the amplitudes are suppressed, showing the effects of condensation and latent heat release on $q$ and $b$ respectively.  In the right panel, the amplitudes of both $q$ and $b$ remain large and the fluctuations are almost exactly out of phase with each other.  This experiment suggests that the proposed latent heat wave damping mechanism is acting to suppress waves in the saturated regions of these atmospheres.

It is interesting to consider whether these moisture modified internal gravity waves might exist in fully saturated atmospheres.  Because the damping of the waves occurs via the negative feedback associated with the latent heat of condensation, and since that coupling between $b$ and $q$ is mediated by $\gamma$ with $\gamma < 1$, it is possible that these waves could be transiently excited in nonlinear simulations.  As a plume transits through the fully saturated atmosphere, the finite amplitude perturbations in $q$ and $b$ that it excites (via $\vec{u}\cdot \vec{\nabla}q$ and $\vec{u}\cdot \vec{\nabla}{b}$) will damp via the negative feedback process explored above.  However, since $\gamma<1$, the cancellation of $b$ by $\gamma q$ is not complete in a single wave period.  We expect that nonlinear plumes may shed transient wakes of internal gravity waves, where the waves damp out over several periods (roughly $1/\gamma$). Indeed, our preliminary non-linear simulations of fully saturated atmospheres have shown evidence of gravity waves driven by moist convective plumes; this was also observed in VPT19. Those authors found (their Figure 16$d-f$) internal gravity waves in the dry subsiding regions of their higher $\Rayleigh$ nonlinear simulations.  This suggests that the mechanism outlined above occurs in both unsaturated atmospheres and fully saturated atmospheres when nonlinear convection can modify the local moisture.  The study of coupled nonlinear convective plumes with possible transient internal gravity waves will be the focus of our next study.

\section{Conclusions}
\label{sec:conclusion}
% The Rainy-Benard system combines the reproducibility, simplicity, and interpretability of Rayleigh-Benard convection with a minimal model of latent heat release due to moisture.
% As such, it offers a promising way to make progress on numerous questions in atmospheric science.
We have presented a detailed treatment of the linear dynamics of the simplified Rainy Benard model for moist convection. We studied convective onset for a variety of parameters of interest in Earth's atmosphere, the spectra of the linear operator for unsaturated atmospheres, and an investigation into the nature of the waves that remain present even when the system has non-oscillatory convective instability present.
For unsaturated atmospheres, we find evidence of linear gravity waves even when the background is unstable.
Unlike in dry Rayleigh-Benard convection, the existence of gravity waves is not precluded by an unstable background. This is true regardless of whether or not the \emph{entire} atmosphere is stable to dry convection. As long as some part of it has a positive buoyancy gradient, we find these waves. We speculated on the existence of similar waves in fully saturated atmospheres from non-linear effects.
This work has precedence in a number of works in the atmospheric literature in which the interaction between conditional instability and gravity waves was considered \citep[e.g.][]{lindzenWaveCISKTropics1974,lindzenBandedConvectiveActivity1976,tulichMultiscaleConvectiveWave2008}.
Here, we have demonstrated that such wave activity may arise from properties of the linear operator even in very simple models of moist convection.

In a future work, we will consider the non-linear evolution of these systems, the scaling of various turbulent quantities with $\Rayleigh$, and the interaction of waves, plumes, and moist convective self-organization.
%\begin{figure}
%    \centering
%    \includegraphics[width=0.4\linewidth]{figures/evp_background_nz256.png} \\
%    \includegraphics[width=0.4\linewidth]{figures/growth_curves_buoyancy_nz256_TSF.png}
%\caption{Non-constant coefficients in the EVP problem.  Growth curves sampled at a series of Rayleigh numbers.}
%\end{figure}

\section*{Acknowledgements}
We would like to thank Geoff Vallis, Steve Tobias, Kasia Nowakowska, and Gregory Dritschel for helpful discussions on moist convection. JSO was supported by DOE EPSCoR grant number DE-SC0024572, while both JSO and BPB were partially supported by NASA SSW 80NSSC19K0026.  The earliest discussions of this work between JSO and BPB occurred at the 2017 Other Worlds Laboratory summer school at University of California Santa Cruz, and we thank that program for their support. Computations were performed on Marvin, a Cray CS500 supercomputer at UNH supported by the NSF MRI program under grant AGS-1919310 and on Premise, a central, shared HPC cluster at the University System of New Hampshire supported by the Research Computing Center and PIs who have contributed compute nodes.

Declaration of Interests. The authors report no conflict of interest.

\appendix
\section{Piecewise solutions for unsaturated atmospheres}
\label{sec:piecewise}
Here, we expand on the discussion in \citet[][VPT19]{vallisSimpleSystemMoist2019a} and give a detailed solution for constructing the unsaturated drizzle solution. This is particularly important in light of the fact that the problem itself is a strongly non-linear boundary value problem but has an important simplification that can be exploited to produce solutions accurately. We draw attention to this because while it is tempting to attempt to solve the full non-linear boundary value problem numerically (using, for example iterative methods), this converges extremely poorly.

In unsaturated atmospheres, equation~\ref{eq:steady_m} continues to hold and $m$ has a linear profile, now with
\begin{align}
P &= \gamma q_0, \label{eq:P_unsaturated}\\
Q & = \beta - 1 + \gamma \Big(\exp{(-\alpha)}-q_0\Big), \label{eq:Q_unsaturated} \\
\end{align}
As in the saturated atmospheres, $m(z) = P + Q z$ is fully determined by the values at the boundaries.  This means that the amplitude and sign of $\Delta m$ can be determined from eqs~\ref{eq:P_unsaturated}--\ref{eq:Q_unsaturated} alone, as can the ideal stability of the atmosphere to moist instabilities.

The profiles of $q(z)$ and $T(z)$ are piecewise functions, with saturated atmosphere solutions for $q_+$, $T_+$ when $z \geq z_c$ and linear solutions for $q_-$, $T_-$ when $z<z_c$, matching at height $z=z_c$ and temperature $T=T_c$:
\begin{align}
    q(z) &=
    \begin{cases}
        q_{-}(z) = q_0+(q_c-q_0)(z/z_c) & z < z_c \\
        q_{+}(z) = \exp(\alpha T_+(z)) & z \geq z_c
    \end{cases} \\
    T(z) &=
    \begin{cases}
        T_{-}(z) = 1 +(T_c-1)(z/z_c) & z < z_c \\
        T_{+}(z) = C(z) - W(\alpha\gamma\exp{(\alpha C(z))})/\alpha & z \geq z_c
    \end{cases}
\end{align}
where
\begin{align}
    C(z) &= b_c + \gamma q_c + \Big((b_2-b_c) + \gamma (q_2-q_c)\Big)\frac{z-z_c}{1-z_c} - \beta z, \\
    b_c &= T_c + \beta z_c, \\
    q_c & = \exp{(\alpha T_c)}.
\end{align}
The buoyancy profile $b$ can be obtained from $b(z) = m(z) - \gamma q(z)$, fully determining the static atmosphere structure.

All that remains is finding the critical values $z_c$ and $T_c$.  This is done, as described in \cite{vallisSimpleSystemMoist2019a}, by requiring $\mathcal{C}^1$ continuity of $q(z)$ and $T(z)$ at $z=z_c$, and solving the resulting nonlinear system via root-finding:
\begin{align}
q_{-}(z) - q_{+}(z) + \tau_1 &= 0, 
 \label{eq:qm_qp}\\
T_{-}(z) - T_{+}(z) + \tau_2 & = 0,\\
\frac{\partial}{\partial z}(q_{-}(z) - q_{+}(z)) & = 0, \\
\frac{\partial}{\partial z}(T_{-}(z) - T_{+}(z)) & = 0, \\
z - z_c &= 0 \label{eq:z_zc},
\end{align}
The nonlinear system of five equations \ref{eq:qm_qp}--\ref{eq:z_zc} can be solved for unknowns $z_c$, $T_c$, along with the slack variables $\tau_1$ and $\tau_2$ (which are zero to machine precision at the root) and $z$ (which equals $z_c$ at the root), using a symbolic tool (e.g., Mathematica or Sympy) or by other methods.

% \begin{figure}
%     \centering
%     \includegraphics[width=\linewidth]{figures/ideal_stability_alpha3.0_Vallis_figure_3.png}
%     \caption{Contours of $\partial_z m$ (blue) and $\mathrm{min}(\partial_z b)$ (red) as functions of $\gamma$ and $\beta$ for systems with $\alpha=3$.  Dashed lines show regions of instability.  The shaded region shows values of $\gamma$ and $\beta$ where the atmosphere is unstable to moist motions ($\partial_z m <0$), but is everywhere stable to dry motions ($\mathrm{min}(\partial_z b > 0$).}
%     \label{fig:ideal_stability_alpha3}
% \end{figure}

\section{A small correction to simulation parameters in VPT19}
\label{sec:VPT19 correction}
In the course of this work, we discovered a small inconsistency in the parameters reported in portions of VPT19, in particular in Figure 4 of that work where they reported critical Rayleigh numbers.  After consulting closely with the authors of that work, we discovered that the code used in computing the nonlinear simulations accidentally adopted
\begin{equation}
    q_{s}^\mathrm{code}(T) = K_2 \exp(\alpha (T_0 + T_1)),
    \label{eq:VPT19 code qs}
\end{equation}
rather than the proper expression
\begin{equation}
    q_{s}(T) = K_2 \exp(\alpha T_1),
    \label{eq:VPT19 correct qs}
\end{equation}
where $K_2$ is a constant.
Equations~\ref{eq:VPT19 code qs} and \ref{eq:VPT19 correct qs} are identical if the temperature at the lower boundary $T_0 = T(z=0) = 0$, as in the analytic sections of VPT19.
The nonlinear simulations of VPT19 instead set $T(z=0)=5.5$ leading to a difference between equations~\ref{eq:VPT19 code qs} and \ref{eq:VPT19 correct qs}.

To correct the results using equation~\ref{eq:VPT19 code qs} from using the full $T=T_0 + T_1$ to the perturbation $T_1$, it is sufficient to adjust the saturated moisture values by a correction factor $G$ with
\begin{equation}
q_s(T) = \frac{q_{s}^\mathrm{code}(T)}{G},
\end{equation}
where $G = K_2 \exp(\alpha T_0)/q_0 \approx 1.542$ (using values of $K_2$, $\alpha$, $T_0$ and $q_0$ from the scripts used in VPT19).

The humidity values $q$ inherit this scaling from $q_s$, and thus equation~\ref{eq:humidity} is left unchanged as all terms scale the same. However, $q$ couples to the buoyancy equation only via $\gamma q$, and so this appears in the dynamics as
\begin{equation}
    \gamma = G \gamma^\mathrm{code},
\end{equation}
implying that the nonlinear simulations of VPT19 are effectively computed at $\gamma=0.293$ rather than the intended value.
By carrying this analysis through to the momentum equation, there is a related correction to the Rayleigh number 
\begin{equation}
    \Rayleigh = \frac{\Rayleigh^\mathrm{code}}{G}.
\end{equation}
With these minor corrections, the critical Rayleigh numbers $\Rac$ reported in VPT19 are in excellent agreement with our calculated values (see figure~\ref{fig:saturated_instability}).  

Finding this small error was only possible by having access to the simulation scripts that the authors of VPT19 used to conduct their work. We emphasize that this highlights the critical importance of making \emph{code} available for inspection.  We sincerely thank Geoff Vallis, Doug Parker, and Steve Tobias for their help and transparency.

%Okay.  It's a bit thick.  But think of how to say this.

\bibliographystyle{jfm}
\bibliography{unsaturated}

\begin{thebibliography}{22}
\expandafter\ifx\csname natexlab\endcsname\relax\def\natexlab#1{#1}\fi
\def\au#1{#1} \def\ed#1{#1} \def\yr#1{#1}\def\at#1{#1}\def\jt#1{\textit{#1}}
  \def\bt#1{#1}\def\bvol#1{\textbf{#1}} \def\vol#1{#1} \def\pg#1{#1}
  \def\publ#1{#1}\def\arxiv#1{#1}\def\org#1{#1}\def\st#1{\textit{#1}}

\bibitem[Bretherton(1987)]{brethertonTheoryNonprecipitatingMoist1987}
{\sc \au{Bretherton, Christopher~S.}} \yr{1987}  \at{A {{Theory}} for
  {{Nonprecipitating Moist Convection}} between {{Two Parallel Plates}}. {{Part
  I}}: {{Thermodynamics}} and ``{{Linear}}'' {{Solutions}}}.  \jt{Journal of
  the Atmospheric Sciences}  \bvol{44}~(14),  \pg{1809--1827}.

\bibitem[Burns {\em et~al.\/}(2024)Burns, Fortunato, Julien \&
  Vasil]{burnsCornerCasesTau2024}
{\sc \au{Burns, Keaton~J.}, \au{Fortunato, Daniel}, \au{Julien, Keith} \&
  \au{Vasil, Geoffrey~M.}} \yr{2024} Corner cases of the tau method:
  Symmetrically imposing boundary conditions on hypercubes,  \arxiv{arXiv:
  2211.17259}.

\bibitem[Burns {\em et~al.\/}(2020)Burns, Vasil, Oishi, Lecoanet \&
  Brown]{burnsDedalusFlexibleFramework2020}
{\sc \au{Burns, Keaton~J.}, \au{Vasil, Geoffrey~M.}, \au{Oishi, Jeffrey~S.},
  \au{Lecoanet, Daniel} \& \au{Brown, Benjamin~P.}} \yr{2020}  \at{Dedalus: {{A
  Flexible Framework}} for {{Numerical Simulations}} with {{Spectral
  Methods}}}.  \jt{Physical Review Research}  \bvol{2}~(2),  \pg{023068},
  \arxiv{arXiv: 1905.10388}.

\bibitem[Chien {\em et~al.\/}(2022)Chien, Pauluis \&
  Almgren]{chienHurricaneLikeVorticesConditionally2022}
{\sc \au{Chien, Mu-Hua}, \au{Pauluis, Olivier~M.} \& \au{Almgren, Ann~S.}}
  \yr{2022}  \at{Hurricane-{{Like Vortices}} in {{Conditionally Unstable Moist
  Convection}}}.  \jt{Journal of Advances in Modeling Earth Systems}
  \bvol{14}~(7),  \pg{e2021MS002846}.

\bibitem[Deng {\em et~al.\/}(2012)Deng, Smith \&
  Majda]{dengTropicalCyclogenesisVertical2012}
{\sc \au{Deng, Qiang}, \au{Smith, Leslie} \& \au{Majda, Andrew}} \yr{2012}
  \at{Tropical cyclogenesis and vertical shear in a moist {{Boussinesq}}
  model}.  \jt{Journal of Fluid Mechanics}  \bvol{706},  \pg{384--412}.

\bibitem[Emanuel(1994)]{emanuelAtmosphericConvection1994}
{\sc \au{Emanuel, K.A.}} \yr{1994} {\em Atmospheric {{Convection}}\/}.
  \publ{Oxford University Press}.

\bibitem[Guichard \& Couvreux(2017)]{guichardShortReviewNumerical2017}
{\sc \au{Guichard, Fran{\c c}oise} \& \au{Couvreux, Fleur}} \yr{2017}  \at{A
  short review of numerical cloud-resolving models}.  \jt{Tellus A: Dynamic
  Meteorology and Oceanography}  \bvol{69}~(1),  \pg{1373578}.

\bibitem[{Hernandez-Duenas} {\em et~al.\/}(2013){Hernandez-Duenas}, Majda,
  Smith \& Stechmann]{hernandez-duenasMinimalModelsPrecipitating2013}
{\sc \au{{Hernandez-Duenas}, Gerardo}, \au{Majda, Andrew~J.}, \au{Smith,
  Leslie~M.} \& \au{Stechmann, Samuel~N.}} \yr{2013}  \at{Minimal models for
  precipitating turbulent convection}.  \jt{Journal of Fluid Mechanics}
  \bvol{717},  \pg{576--611}.

\bibitem[{Hernandez-Duenas} {\em et~al.\/}(2015){Hernandez-Duenas}, Smith \&
  Stechmann]{hernandez-duenasStabilityInstabilityCriteria2015}
{\sc \au{{Hernandez-Duenas}, Gerardo}, \au{Smith, Leslie~M.} \& \au{Stechmann,
  Samuel~N.}} \yr{2015}  \at{Stability and {{Instability Criteria}} for
  {{Idealized Precipitating Hydrodynamics}}}.  \jt{Journal of the Atmospheric
  Sciences}  \bvol{72}~(6),  \pg{2379--2393}.

\bibitem[Lindzen(1974)]{lindzenWaveCISKTropics1974}
{\sc \au{Lindzen, Richard~S.}} \yr{1974}  \at{Wave-{{CISK}} in the
  {{Tropics}}}.  \jt{Journal of the Atmospheric Sciences}  \bvol{31}~(1),
  \pg{156--179}.

\bibitem[Lindzen \& Tung(1976)]{lindzenBandedConvectiveActivity1976}
{\sc \au{Lindzen, R.~S.} \& \au{Tung, K.-K.}} \yr{1976}  \at{Banded
  {{Convective Activity}} and {{Ducted Gravity Waves}}}.  \jt{Monthly Weather
  Review}  \bvol{104}~(12),  \pg{1602--1617}.

\bibitem[Oishi {\em et~al.\/}(2021)Oishi, Burns, Clark, Anders, Brown, Vasil \&
  Lecoanet]{oishiEigentoolsPythonPackage2021a}
{\sc \au{Oishi, Jeffrey}, \au{Burns, Keaton}, \au{Clark, S.}, \au{Anders,
  Evan}, \au{Brown, Benjamin}, \au{Vasil, Geoffrey} \& \au{Lecoanet, Daniel}}
  \yr{2021}  \at{Eigentools: {{A Python}} package for studying differential
  eigenvalue problems with an emphasis on robustness}.  \jt{Journal of Open
  Source Software}  \bvol{6}~(62),  \pg{3079}.

\bibitem[Pauluis \& Schumacher(2010)]{pauluisIdealizedMoistRayleighBenard2010}
{\sc \au{Pauluis, Olivier} \& \au{Schumacher, J{\"o}rg}} \yr{2010}
  \at{Idealized moist {{Rayleigh-Benard}} convection with piecewise linear
  equation of state}.  \jt{Communications in Mathematical Sciences}
  \bvol{8}~(1),  \pg{295--319}.

\bibitem[Pauluis \&
  Schumacher(2011)]{pauluisSelfaggregationCloudsConditionally2011}
{\sc \au{Pauluis, Olivier} \& \au{Schumacher, J{\"o}rg}} \yr{2011}
  \at{Self-aggregation of clouds in conditionally unstable moist convection}.
  \jt{Proceedings of the National Academy of Sciences}  \bvol{108}~(31),
  \pg{12623--12628}.

\bibitem[Pauluis \& Schumacher(2013)]{pauluisRadiationImpactsConditionally2013}
{\sc \au{Pauluis, Olivier} \& \au{Schumacher, J{\"o}rg}} \yr{2013}
  \at{Radiation {{Impacts}} on {{Conditionally Unstable Moist Convection}}}.
  \jt{Journal of the Atmospheric Sciences} .

\bibitem[Rohatgi(????)]{WebPlotDigitizer}
{\sc \au{Rohatgi, Ankit}} \yr{????} Webplotdigitizer.

\bibitem[Schumacher \& Pauluis(2010)]{schumacherBuoyancyStatisticsMoist2010}
{\sc \au{Schumacher, J{\"o}rg} \& \au{Pauluis, Olivier}} \yr{2010}
  \at{Buoyancy statistics in moist turbulent {{Rayleigh}}--{{B{\'e}nard}}
  convection}.  \jt{Journal of Fluid Mechanics}  \bvol{648},  \pg{509--519}.

\bibitem[Spiegel \&
  Veronis(1960)]{spiegelBoussinesqApproximationCompressible1960}
{\sc \au{Spiegel, E.~A.} \& \au{Veronis, G.}} \yr{1960}  \at{On the
  {{Boussinesq Approximation}} for a {{Compressible Fluid}}.}  \jt{The
  Astrophysical Journal}  \bvol{131},  \pg{442}.

\bibitem[Stevens(2005)]{stevensATMOSPHERICMOISTCONVECTION2005}
{\sc \au{Stevens, Bjorn}} \yr{2005}  \at{{{ATMOSPHERIC MOIST CONVECTION}}}.
  \jt{Annual Review of Earth and Planetary Sciences}  \bvol{33}~(1),
  \pg{605--643}.

\bibitem[Tulich \& Mapes(2008)]{tulichMultiscaleConvectiveWave2008}
{\sc \au{Tulich, Stefan~N.} \& \au{Mapes, Brian~E.}} \yr{2008}  \at{Multiscale
  {{Convective Wave Disturbances}} in the {{Tropics}}: {{Insights}} from a
  {{Two-Dimensional Cloud-Resolving Model}}}.  \jt{Journal of the Atmospheric
  Sciences}  \bvol{65}~(1),  \pg{140--155}.

\bibitem[Turner(1973)]{Turner_1973}
{\sc \au{Turner, J.~S.}} \yr{1973} {\em Buoyancy Effects in Fluids\/}.
  \publ{Cambridge University Press}.

\bibitem[Vallis {\em et~al.\/}(2019)Vallis, Parker \&
  Tobias]{vallisSimpleSystemMoist2019a}
{\sc \au{Vallis, Geoffrey~K.}, \au{Parker, Douglas~J.} \& \au{Tobias,
  Steven~M.}} \yr{2019}  \at{A simple system for moist convection: The
  {{Rainy}}--{{B{\'e}nard}} model}.  \jt{Journal of Fluid Mechanics}
  \bvol{862},  \pg{162--199}.

\end{thebibliography}

\end{document}